\RequirePackage{fix-cm}
\documentclass[smallcondensed]{svjour3}     
\smartqed  
\usepackage{graphicx,verbatim}
\usepackage{float,hyperref}
\usepackage{amsmath, amssymb, amsfonts}
\usepackage{appendix}
\newcommand{\norm}[1]{\left\lVert #1 \right\rVert}

\newcommand{\LyapunovDim}{p}

\newcommand{\curve}{{\mathcal C}}
\newcommand{\curvepi}{{\mathcal C}_{x,i}}
\newcommand{\tildeW}{\tilde{W}}
\begin{document}

\title{An ergodic-averaging method to differentiate covariant Lyapunov vectors \thanks{This work was supported by Air Force Office of Scientific Research Grant No. FA8650-19-C-2207.}
}
\subtitle{Computing the curvature of one-dimensional unstable manifolds of strange attractors}


\author{Nisha Chandramoorthy         \and
        Qiqi Wang 
}


\institute{N. Chandramoorthy \at
              Department of Mechanical Engineering and Center for Computational Science and Engineering \\
			  Massachusetts Institute of Technology \\
              77 Massachusetts Avenue, Cambridge MA 02139 \\
              \email{nishac@mit.edu}           
           \and
           Q. Wang \at
              Department of Aeronautics and Astronautics and Center for Computational Science and Engineering \\
			  Massachusetts Institute of Technology \\
              77 Massachusetts Avenue, Cambridge MA 02139 \\
              \email{qiqi@mit.edu}
}

\date{Submitted version}

\maketitle

\begin{abstract}
Covariant Lyapunov vectors or CLVs span the expanding and contracting 
directions of perturbations along trajectories in a chaotic dynamical 
system. Due to efficient algorithms to 
compute them that only utilize trajectory information, they have been 
widely applied across scientific disciplines, principally for sensitivity 
analysis and predictions under uncertainty. In this paper, we develop a  
numerical method to compute the directional derivatives of the first CLV along its own 
direction; the norm of this derivative is also the curvature of one-dimensional unstable manifolds. Similar to the computation of CLVs, the present method for their 
derivatives is iterative and analogously uses the second-order derivative 
of the chaotic map along trajectories, in addition to the Jacobian.
We validate the new method on a super-contracting Smale-Williams Solenoid attractor. We also demonstrate the algorithm on several other examples including smoothly perturbed Arnold Cat maps, and the Lorenz'63 attractor, obtaining visualizations of the curvature of each attractor. Furthermore, we reveal a fundamental connection of the derivation of the CLV self-derivative computation with an efficient computation of linear response of chaotic systems.
\end{abstract}

\section{Introduction}
\label{intro}
{\em Linear response} refers to the linear change in the long-term or statistical behavior of a dynamical system, as a result of a small parameter perturbation. In chaotic systems, a linear response formula was developed by Ruelle \cite{ruelle}\cite{ruelle1}, which is rigorously proved for {\em uniformly hyperbolic} systems, the simplest setting in which a chaotic attractor can occur.  Linear response has been observed in practical chaotic systems wherein dissipative dynamics dominate \cite{wormell}\cite{angxiu-jfm}\cite{patrick-fluid-1}\cite{nisha-shadowing}\cite{francisco}. A unique ergodic stationary physical probability distribution, known as an SRB measure \cite{srb}, is achieved on uniformly hyperbolic attractors. Linear response gives us a quantitative estimate of the derivative of the SRB measure with respect to system parameters, using information only from the unperturbed system. 

This statistical derivative can enable typical applications of sensitivity analysis, such as uncertainty quantification, design, optimization and control problems in chaotic systems. These applications are currently limited in chaotic systems because the computation of linear response, through Ruelle's theoretical formula, remains a challenging problem. Some new numerical methods are being actively developed as of this writing (\cite{nisha-s3}\cite{angxiu}) in which Ruelle's formula is transformed into a well-conditioned ergodic-averaging computation; other promising methods include shadowing-based methods \cite{qiqi-shadowing}\cite{angxiu-nilss}, and approximate evaluations of Ruelle's response using fluctuation-dissipation theorems extended to SRB-type measures \cite{majda1}\cite{lucarini-linear-response}\cite{majda}. 

In this work, we develop a numerical method for derivatives on the unstable manifold of certain quantities fundamental to linear response. These derivatives are needed for an efficient computation of a regularized version of Ruelle's formula. Focusing on one-dimensional unstable manifolds, the proposed numerical method gives the derivative of the unstable Covariant Lyapunov Vector (CLV) \cite{kuptsov} along its own direction. As a byproduct, we obtain the unstable derivative of the local expansion factor of the unstable CLV, and this quantity appears in the computation of linear response.

We expect the CLV self-derivatives computed in this paper, which describe the curvature of the attractor manifold, to be applicable beyond linear response. CLVs are specific bases for tangent spaces along a trajectory, characterized by Lyapunov 
exponents. Ginelli {\emph et al.}'s \cite{ginelli}  efficient 
algorithm to compute CLVs has led to several applications of Lyapunov analysis in engineering, 
in both deterministic and stochastic chaotic 
systems. 
These applications include uncertainty quantification, data assimilation and forecasting, across a range of disciplines such as numerical weather prediction and aerospace engineering (\cite{beeson}, \cite{angxiu-jfm}, \cite{lucarini_climate}, \cite{francisco}; see \cite{ginelli-clv-review} for a survey of applications of Lyapunov analysis).

The numerical method we develop in this work for the directional derivatives of CLVs in their respective directions is henceforth known as the {\em differential CLV method}. We shall refer to these derivatives as \emph{CLV self-derivatives}. In the case of a one-dimensional unstable manifold, the CLV corresponding to the 
largest Lyapunov exponent is the unit tangent vector field along the unstable manifold. The norm of this CLV self-derivative is hence also the curvature of the unstable manifold. 

The connection we reveal with linear response is via a byproduct of the differential CLV method: the unstable derivative of the local expansion factors of the unstable CLVs. This derivative is a key ingredient in an iterative computation of a fundamental quantity intimately connected to linear response. This quantity, which we refer to as the logarithmic {\em density gradient}, indicates how the SRB measure changes along unstable manifolds in the attractor. More precisely, in the case of one-dimensional unstable manifolds, which is the focus of this paper, this quantity is the unstable derivative of the logarithm of the SRB density on the unstable manifold. This connection shows one potential application of the recursive method developed in this paper: the computation of linear response in chaotic systems.

The outline of the subsequent sections is as follows. In section \ref{sec:background}, 
we briefly summarize the theory of CLVs and establish the setting we derive our results in: uniformly hyperbolic attractors. The \emph{differential CLV method} is derived in section \ref{sec:algorithm}; while the main steps are in section 
\ref{sec:mainDerivation}, notational setup and the intuition for the steps are developed in the prior subsections. We validate the method using a super-contracting Solenoid map in section \ref{sec:solenoid}. Further numerical experiments demonstrating the method on the Lorenz'63 attractor, a volume-preserving perturbed Cat map, a dissipative perturbed Cat map, and the H\'enon map are in sections \ref{sec:lorenz}, \ref{sec:pcm}, \ref{sec:dpcm} and 
\ref{sec:henon} respectively. 
The implication of the method for the computation of linear response is discussed in section \ref{sec:linearResponse}.
We summarize our results and conclude in section \ref{sec:conclusions}. 

\section{Problem setup, definitions and review of Covariant Lyapunov Vectors}
\label{sec:background}
The dynamical 
system studied in this paper is the iterative application of a smooth ($C^3$) self-map $\varphi:\mathbb{M} \to \mathbb{M}$ of a domain $\mathbb{M}$, which is a compact subset of $\mathbb{R}^m$. We write $\varphi^n$ to denote an $n$-time composition of $\varphi$; that is, $\varphi^n = \varphi\circ \varphi^{n-1}$, $n \in \mathbb{Z}^+$, where $\varphi^0$ is the identity function on $\mathbb{R}^m$. The iterates under $\varphi$, or the points along orbits of the dynamical system, are represented using 
the following subscript notation: if $x \in \mathbb{M}$,  
$x_n := \varphi^n x$; $x_0$ is simply written as $x$, which we use to denote an arbitrary phase point. A similar notation is also adopted for scalar or vector-valued
functions or observables. If $f$ is an observable,
$f_n := f\circ \varphi^n.$ The derivative with respect to 
the state is denoted as $d$ and the partial derivative operators,
with respect to the Euclidean coordinate functions $x_1,x_2,\cdots,x_m$ are written as 
$\partial_1,\partial_2,\cdots,\partial_m,$ respectively. For instance, if $f:\mathbb{M}\to \mathbb{R}$ 
is a scalar-valued observable, the derivative $df$ evaluated at $x$ is given by 
$df(x) =[\partial_1f(x),\cdots,\partial_mf(x)]^T.$ Using the notation introduced, an application of the chain rule would be as follows:
$$(df_n)^T = ((df)_n)^T \; d\varphi^n.$$ Finally, we assume the existence of an ergodic, physical, invariant measure for $\varphi$, 
known as the SRB measure and denoted $\mu$. As a result, ergodic (Birkhoff) averages of observables in $L^1(\mu)$ converge to their expectations with respect to $\mu$: $\lim_{N\to\infty} (1/N)\sum_{n=0}^{N-1} f_n(x) = \langle f, \mu\rangle$\ for Lebesgue-a.e. $x \in \mathbb{M}.$ 
Note that such a measure is guaranteed to exist \cite{srb} in the uniformly hyperbolic setting, which we discuss in section \ref{sec:uniformHyperbolicity}. 
\subsection{Tangent dynamics}
In order to introduce covariant Lyapunov vectors (CLVs), whose 
derivatives are the subject of this paper, we briefly discuss 
the asymptotic behavior of tangent dynamics in chaotic systems. We refer to 
as tangent dynamics the linear evolution of perturbations under the 
Jacobian matrix, $d\varphi$. The Jacobian matrix evaluated at an $x \in \mathbb{M}$ is denoted $d\varphi_x$. 
Denoting the tangent space at $x$ as 
$T_x \mathbb{M}$, $d\varphi^n_x$ is a map from $T_x\mathbb{M}$ to 
$T_{x_n}\mathbb{M}$. Given a tangent vector $v_0 \in T_x\mathbb{M}$, 
we denote its iterate under the tangent dynamics at time $n$ as $v_n 
\in T_{x_n}\mathbb{M}.$ That is, $v_n = d\varphi^n_x v_0.$ Intuitively, 
if a perturbation of norm ${\cal O}(\epsilon)$ is applied at $x$ along $v_0$, up to 
first order in $\epsilon$, the deviation from the original orbit starting at $x$, after time $n$, is along $v_n$. In other words, 
\begin{align}
    \label{eqn:tangentEquation}
    v_n = \lim_{\epsilon \to 0} \dfrac{\varphi^n(x + \epsilon v_0) - x_n}{\epsilon}
    = d\varphi^n_x v_0.
\end{align} 
In practice, the above equation for the tangent dynamics is solved iteratively, along a reference orbit $\left\{x_n\right\}$, since using the chain rule, $d\varphi^n_x = d\varphi_{x_{n-1}}\cdots d\varphi_x,$
and hence $v_{n+1} = d\varphi_{x_n} v_n.$ A classical result in nonlinear dynamics, known as the Oseledets multiplicative ergodic theorem (OMET) \cite{arnold} deals with the asymptotic behavior of $v_n$ as $n\to\infty$, in ergodic systems. The OMET implies the following: at $\mu$-a.e. $x \in \mathbb{M}$, the tangent space splits as a direct sum, $T_x \mathbb{M} = \oplus_{i=1}^{\LyapunovDim} E^i_x$, $\LyapunovDim \leq m$, where $E^i_x$ are $d\varphi$-invariant 
subspaces in the sense that $d\varphi_x E^i_x = E^i_{\varphi x}$. This splitting is based on the asymptotic, exponential growth/decay rates of
tangent dynamics in the subspaces $E^i_x$. More precisely, for $\mu$-a.e. $x \in \mathbb{M}$, if $v^i_0 
\in E^i_x$, its norm under the tangent dynamics grows/decays exponentially
at a rate that converges to a constant. The limits
\begin{align}
		\lambda_{x,i} := \lim_{n\to\infty}\dfrac{1}{n}
		\log\left(\dfrac{\norm{v^i_n}}{\norm{v^i_0}}\right),
\end{align}
$1\leq i\leq \LyapunovDim$, are known as the Lyapunov exponents (LEs). Since $\varphi$ is an ergodic map with respect to $\mu$, the LEs are 
constants independent of $x$, for $\mu$-a.e. $x$; we denote the LEs $\lambda_i$, in descending order as $\lambda_1 \ge \lambda_2 \ge \cdots \ge \lambda_{\LyapunovDim}.$ In our setting, $\varphi$ is a chaotic map, which means that $\lambda_1 > 0$. Let $d_u$ be the number of positive LEs, and $d_s = \LyapunovDim - d_u$ be the number of negative LEs. Then, $E^u_x := \oplus_{i=1}^{d_u} E^i_x$ is called the unstable subspace of
$T_x \mathbb{M}.$ In other words, 
the unstable subspace $E^u_x$ is the set of tangent 
vectors that asymptotically decay exponentially in norm under tangent dynamics backward in time; by definition, the unstable subspaces at points on a chaotic orbit are non-empty. This sensitivity to perturbations is the so-called {\em butterfly effect} that defines chaotic systems. 

Similarly,
the set of tangent vectors that asymptotically decay exponentially in norm under the tangent dynamics, 
make up the stable subspace, denoted $E^s_x := \oplus_{i=d_u +1}^\LyapunovDim E^i_x = T_x\mathbb{M} \backslash E^u_x$. If each $E^i$ is one-dimensional and $p=m$, the covariant Lyapunov vectors or CLVs, denoted as $V^i$ in this paper, are unit vector fields along $E^i.$ That is, CLVs satisfy the following properties at $\mu$-a.e. $x$:
\begin{itemize}
    \item The covariance property: \begin{align}
					d\varphi_x V^i_x \in E^i_{x_1}. 
    \end{align}
				Since by definition $V^i_{x}$ is a unit vector, we introduce 
				a scalar function $z_{\cdot, i}:\mathbb{M}\to\mathbb{R}^+$ defined as 
				$z_{x,i} := \norm{d\varphi_x V^i_x},$
    to indicate the local stretching or contraction factor of 
    the $i$th CLV. Hence, the covariance property of the 
    $i$th CLV can be expressed as
    \begin{align}
     \label{eqn:covariance}
			d\varphi_x V_{x,i} = z_{x,i} V^i_{x_1}.
    \end{align}
    \item The $i$th CLV grows/decays asymptotically on an exponential scale, at the rate $\lambda_i$, and, in addition, is invariant 
    under time-reversal:
    \begin{align}
        \label{eqn:LyapunovExponents}
        \lambda_i := \lim_{n\to\pm\infty}\frac{1}{n} \log
			\norm{d\varphi^n_x V^i_x}.
    \end{align}
\end{itemize}
\subsection{Uniform hyperbolicity}
\label{sec:uniformHyperbolicity}
We consider an idealized class of chaotic systems known as uniformly hyperbolic systems, which are characterized by uniform expansions and contractions of tangent vectors. In uniformly hyperbolic systems, there exist constants $c > 0$ and $\lambda \in (0,1)$ such that, at every point $x \in \mathbb{M}$, i) every stable tangent vector $v \in E^s_x$ satisfies: $\norm{d\varphi^n_x v} \leq c\; \lambda^n \norm{v}$, and 
ii) every unstable tangent vector 
$v \in E^u_x$ satisfies: $\norm{d\varphi^{-n}_x v} 
\leq c \;\lambda^n\norm{v}$, for all $n \in \mathbb{N}$. As a result, in these systems, there exists an upper (lower) bound that is independent of the base point $x$, on the slowest contracting (stretching) factors among  $z_{x,i}$. In particular, defining $C := c \lambda$,
we have $z_{x,i} \geq (1/C)$, $1\leq i\leq d_u$ and 
$z_{x,i} \leq C$, $d_u+1\leq i\leq d.$ 
From the definition of the LEs (Eq. \ref{eqn:LyapunovExponents}), it is also clear 
that they 
are the ergodic (Birkhoff) averages of the stretching/contraction
factors: 
\begin{align}
    \label{eqn:ergodicAverageOfzi}
		\langle \log z_{\cdot, i},\mu\rangle := \lim_{N\to\infty} 
		\frac{1}{N} \sum_{n=0}^{N-1} \log{z_{x_n,i}} = 
\lambda_i, \;\; x \in \mathbb{M}\;\; \mu-{\rm a.e}. 
\end{align}

\subsection{Examples}
\label{sec:examples}
A simple example of a uniformly hyperbolic 
system is Arnold's Cat map, a smooth self-map of the 
surface of the torus ($\mathbb{T}^2 \equiv \mathbb{R}^2/\mathbb{Z}^2$):
\begin{align}
		\varphi([{\rm x}_1,{\rm x}_2]^T) = \begin{bmatrix}
        2 & 1 \\
        1 & 1 
    \end{bmatrix}
    \begin{bmatrix}
			{\rm x}_1 \\
			{\rm x}_2
    \end{bmatrix}  \; {\rm mod}\; 1. 
\end{align}
This is a linear hyperbolic 
system, i.e., the Jacobian matrix of the map is a constant 
in phase space and has eigenvalues other than 1. In this simple example, the CLVs and the stretching/contracting 
factors, are also independent of the phase point. The logarithm of the eigenvalues of the constant Jacobian matrix,
are the LEs of this map: $\lambda_1 = \log|(3 + \sqrt{5})/2|$ and $\lambda_2 = \log|(3 - \sqrt{5})/2|.$ It is also clear that $E^1 = E^u$ and $E^2 = E^s$ are one-dimensional subspaces 
spanned by $V^1$ and $V^2$, the eigenvectors of 
the Jacobian matrix at eigenvalues of $e^{\lambda_1}$
 and $e^{\lambda_2}$ respectively. Moreover, $z_1$ and $z_2$ are also constant 
 on $\mathbb{R}^2/\mathbb{Z}^2$: $z_1=e^{\lambda_1}$, and $z_2 = e^{\lambda_2}$. Further, the SRB measure for this map is the Lebesgue measure on $\mathbb{R}^2/\mathbb{Z}^2.$ 
 
 Since the Jacobian 
 matrix is symmetric, the CLVs $V^1$ and $V^2$ are everywhere 
 orthogonal to each other, but it is worth noting that 
 this is a special case. In a generic  
 uniformly hyperbolic system, it is only true that 
 the angle between the CLVs is uniformly bounded away from zero. The perturbed Cat maps treated 
 later have additive perturbations to the Cat map 
 above that are smooth functions on the torus. Two types of smooth perturbations are considered later, both designed to produce non-uniform behavior of the CLVs. Both perturbed Cat maps are still uniformly 
 hyperbolic, and differ in 
 whether or not the resulting maps are area-preserving,
in order to represent the two distinct cases of conservative (symplectic) and dissipative chaos.
 
\subsection{Lack of differentiability of $E^u$ and $E^s$}
\label{sec:irregularity}
On hyperbolic sets, it is known that $E^u$ and $E^s$ are H\"older-continuous functions of phase space, 
in a sense clarified in Appendix section \ref{sec:AppxContinuity}. When the H\"older exponent 
$\beta$, from Appendix section \ref{sec:AppxContinuity}, equals 1, we have Lipschitz continuity, but this is indeed rare. Several examples (see \cite{hasselblatt_prevalence_1999} and references therein) have been constructed in which $\beta$ is made to be arbitrarily small at almost all phase points, even in $C^\infty$ maps. In rare cases, 
$E^u$ and $E^s$ are continuously differentiable 
when a certain bunching condition 
(\cite{hasselblatt_prevalence_1999}, or 
section 19.1 of \cite{katok}) is satisfied 
by the LEs. 

Revisiting the examples, the perturbed Cat maps 
discussed above belong to the rare category of 
maps with continuously differentiable stable/unstable subspaces. In fact, it can be shown that all uniformly hyperbolic maps on 
compact sets of dimension 2 belong to this category (see 
Corollary 19.1.11 of \cite{katok}). While it would 
be typical of a higher-dimensional map, even when 
uniformly hyperbolic, to show non-smoothness of 
the stable and unstable subspaces, we have chosen
to work with two-dimensional examples in this paper 
for easy visualization of the subspaces, which 
are lines in these maps. 

\subsection{Derivatives of CLVs in their own directions}
While the CLVs may lack differentiability on $\mathbb{M}$,
they have directional derivatives in their own 
directions. In fact, it can be shown that these directional 
derivatives, which we refer to here as CLV self-derivatives, are themselves H\"older continuous 
with the same exponent $\beta$ (see Remark in the proof of Theorem 19.1.6 of \cite{katok}). To wit, in two-dimensional uniformly hyperbolic systems, examples of which are considered in this paper, both partial derivatives (along coordinate directions) of the CLVs exist, and hence the CLVs have directional derivatives in all directions. The purpose of this paper, however, is to 
numerically compute directional derivatives of 
CLVs along their respective directions in a general 
uniformly hyperbolic system, regardless of their 
differentiability in phase space. Thus, we compute the CLV self-derivatives, without using the partial derivatives along coordinate directions, which may not exist. The CLV self-derivatives are denoted by $W^i_x \in
T_x T_x \mathbb{M} \equiv \mathbb{R}^m$. They are defined using curves  
$\curve_{x,i}:[-\epsilon_x,\epsilon_x]\to \mathbb{M}$ with the properties:  
i) $\curve_{x,i}(0) = x$, ii) $\curve_{x,i}'(t) = 
V^i(\curve_{x,i}(t)), \;\forall\;t \in [-\epsilon_x,
\epsilon_x]$, as
\begin{align}
\label{eqn:Widefn}
		W^i_x := \lim_{t\to 0} \dfrac{V^i_{\curve_{x,i}(t)} - V^i_x}{t}. 
\end{align}
For example, in the case of a 1-dimensional unstable manifold, 
the curve $\curve_{x,1}$, coincides with a 
local unstable manifold at $x.$ Further discussion on the definition 
of $W^i$ based on these curves, is postponed until section \ref{subsec:coordinateNotation}. Here we explain the existence 
of these curves. The vector fields $V^i$, $1\leq i\leq d_u$ are infinitely smooth on an open set in a local unstable manifold, and likewise, $V^i$, for $d_u +1 \leq i \leq d$ are infinitely smooth on an open set in a local stable manifold. As a result, due to the existence and uniqueness theorem, the flow of vector field 
$V^i$, denoted by the curve $\curve_{\cdot, i}$ exists and is uniquely defined, for some $\epsilon_\cdot > 0,$ justifying the definition in 
Eq. \ref{eqn:Widefn}. 

Given
$T_x \mathbb{M}, T_x T_x \mathbb{M} \equiv \mathbb{R}^m$, we 
write all vectors in these spaces in Euclidean coordinates.
The output of the numerical method to be developed, $W^i$, are
$d$-dimensional vector fields consisting of component-wise 
directional derivatives of $V^i.$ 

\subsection{Computations along trajectories}
Before we delve into the differential CLV method, we note that $W^i$, being self-derivatives of CLVs, are naturally defined along trajectories, just like the CLVs. Thus, we seek a trajectory-based iterative procedure to compute them. We assume as input to the method the map, its Jacobian and second-order derivative, all computed along a long, $\mu$-typical trajectory. The CLVs that need to be differentiated are also assumed as input, along the trajectory. To compute the CLVs, a standard algorithm such as Ginelli {\emph et al.}'s algorithm \cite{ginelli} can be used. This is an iterative procedure involving repeated QR factorizations of nearby subspaces to the one that is spanned by the required CLVs. For Ginelli {\emph et al.'s} algorithm, the reader is referred to \cite{ginelli} and \cite{florian} 
for its convergence with respect to trajectory length; for other algorithms that involve LU factorizations instead of QR, we refer to \cite{kuptsov}. 

Besides using the computed CLVs as input, the differential CLV method we develop here for $W^i$ does not follow Ginelli {\emph et al.}'s or other algorithms for the computation of CLVs, primarily because the vector fields $W^i$ do not satisfy the covariance property. But the method resembles the latter algorithms in being iterative and trajectory-based. One advantage of trajectory-based computation is that we  exploit for fast convergence (this aspect again being similar to the CLV computation algorithms) the hyperbolic splitting 
of the tangent space. This will be clear at the end of the next section in which we give a step-by-step derivation. 

\section{An algorithm to compute the directional derivatives of CLVs in their own directions}
\label{sec:algorithm}
In this section, we derive a numerical method to determine the quantity of interest, $W^i,$ which is defined in Eq. \ref{eqn:Widefn}. 
In particular, fixing a reference trajectory $x,x_1,\cdots,$
we develop an iterative scheme that converges asymptotically to vectors $W^i_{n} := W^i_{x_n}$, under certain conditions (Appendix \ref{sec:AppxConvergence}), starting 
from an arbitrary guess for $W^i_0 := W^i_x \in \mathbb{R}^m.$ The derivation results in the 
following iteration, valid for 
$1\leq i \leq d_u, n\in\mathbb{Z}^+$, and guaranteed to converge when $i=1$:
\begin{align}
\label{eqn:iterationFirst}
		W^i_{n+1} = \Big(I - V^i_{n+1} (V^i_{n+1})^T \Big) \dfrac{d^2\varphi(x_n): V^i_n \: V^i_n + d\varphi(x_n) W^i_n}{z_{n,i}^2}. 
\end{align}
The iteration mainly uses the 
chain rule and the covariance property of $V^i$, in 
a convenient set of 
coordinate systems centered along each 
$\mu$-typical trajectory. These trajectory-based coordinates 
help us uncover each term on the right hand side of Eq. \ref{eqn:iterationFirst}. 

\subsection{Change of coordinates and associated notation}
\label{subsec:coordinateNotation}
Fix a $\mu$-typical point $x \in \mathbb{M}$, and 
consider again the curves $\curvepi$, $1\leq i\leq d,$
which were introduced to define $W^i$ in Eq. \ref{eqn:Widefn}. To reiterate, the curves $\curvepi:[-\epsilon_x, \epsilon_x] \to \mathbb{M}$ are such that 
i) $\curvepi(0) = x$ and ii) $\curvepi'(t) = V^i(\curvepi(t)),$
for all $t \in [-\epsilon_x,\epsilon_x].$ There exists 
a measurable function $x\to \epsilon_x$ that defines 
the extent of the curves so that such a coordinate 
change, from $[-\epsilon_x,\epsilon_x]^m$ to a neighborhood 
of $x$, exists and is additionally differentiable. This follows from an assertion 
proved in standard stable-unstable manifold theory: a closed 
$\epsilon_x$ Euclidean ball around the origin in $\mathbb{R}^{d_u}$
($\mathbb{R}^{d_s}$) has an embedding into 
a local unstable (stable) manifold at $x$. These pointwise coordinate 
systems are referred to as Lyapunov charts
or adapted coordinates in the theoretical literature (\cite{katok} Ch. 6, \cite{ledrappier-young}). 

In writing Eq. \ref{eqn:Widefn}, we made a particular choice of adapted coordinates. We chose coordinate functions that are adapted specifically to the CLVs, as opposed to any other basis of $T_x \mathbb{M}$, in the following sense. At each $x$,
the image of the $i$th Euclidean basis vector $e_i$, under the differential of the coordinate change, 
is $V^i$. More intuitively, we have chosen adapted coordinates 
such that the $i$th Euclidean coordinate axis corresponds, under 
these coordinate changes, to points that  
are perturbations along $V^i$. Thus, our quantity of interest, can be written, by definition of CLV-adapted coordinates, as 
\begin{align}
    \label{eqn:WiinNewCoordinates}
    W^i_x = \frac{d}{dt} (V^i\circ\curvepi)(0).
\end{align}
\subsection{The map in adapted coordinates}
Now we introduce the transformation induced by $\varphi:\mathbb{M} \to \mathbb{M}$ on the CLV-adapted coordinates on $\mathbb{R}^m$. To do that, we fix an $i\leq d_u$ and focus on the relationship between
the curves $\curve_{x_1, i}:[-\epsilon_{x_1},\epsilon_{x_1}]\to\mathbb{M}$ 
and $\varphi\circ \curve_{x,i}:[-\epsilon_x,\epsilon_x]\to\mathbb{M}.$
Define $f_{x,i} := (\curve_{x_1,i})^{-1} \circ \varphi \circ \curvepi,$
noting that this definition makes sense at a point $t \in [-\epsilon_x, \epsilon_x]$ whenever $\varphi(\curvepi(t))$ lies in the image of $\curve_{x_1,i}.$ The function, $x \to \epsilon_x$, which determines the size of the local unstable manifold at each $x$, can be chosen such that orbits of 
\begin{align}
\label{eqn:conditionOnCoordinateSystem}
		f_{x,i}^{-n} := f_{x_n,i}^{-1} \circ \cdots \circ f_{x_1,i}^{-1}\circ f_{x,i}^{-1}, 
n\in \mathbb{Z}^+
\end{align}
are well-defined at almost every $x$, for $1\leq i\leq d_u,$ within local unstable manifolds centered along the backward $\varphi$-orbit. Clearly, 0 is a fixed point 
of $f_{x,i}^n$ for all $n \in \mathbb{Z}$, and corresponds to 
the $\varphi$-orbit $\cdots, x_{-1}, x,x_1,x_2,\cdots,.$ Intuitively, 
if an orbit of $f_{x,i}$ excluding the fixed point, say 
$\left\{t_n := f_{x, i}^n(t)\right\}$, exists, it 
means that $\curve_{x_n,i}(t_n)$ lies in sufficiently small local unstable manifolds of $x_n$, at each $n.$ The sizes of the local unstable manifolds can be controlled in order for such orbits to be well-defined (in particular, see Lemma 2.2.2 of \cite{ledrappier-young}). 

To summarize, we make a specific choice of $x\to \epsilon_x$ such 
that the curves $\curve_{x_n, i}$ at each $n$ lie inside a local 
unstable manifold at $x_n$, and are tangent to $V^i_n := V^i_{x_n}.$
This allows us to obtain 
expressions for the derivative of a CLV $V^i_{n+1}$ with respect 
to $V^i_n,$ which will in turn enter into the computation of $W^i.$ 
In particular, using CLV-adapted coordinates, a suitable map $x\to \epsilon_x$, as described above, and the definition of $f_{x,i}$,
\begin{align}
\label{eqn:dfpi}
		(df_{x,i}/dt)(0) = z_{x,i}. 
\end{align}
Now we usefully relate the iterates through $\varphi$ of the differential operator on $\mathbb{M}$ corresponding to the vector field $V^i$, and its analog on $\mathbb{R}$: $d/dt$, along the 
trajectory lying in the unstable manifold of $x_n$. In particular, for the function $V^i$, 
when combined with Eq. \ref{eqn:WiinNewCoordinates}, and Eq. \ref{eqn:dfpi}, 
\begin{align}
    \notag
    \frac{d \left(V^i \circ \varphi \circ \curvepi\right)}{dt} (0) &=
    \frac{d \left(V^i \circ \curve_{x_1,i} \circ f_{x,i}\right)}{dt} (0) \\
    \label{eqn:Wip1}
		&= z_{x,i} \; W^i_{x_1}.
\end{align}
\subsection{Computation of unstable CLV self-derivatives}
\label{sec:mainDerivation}
Starting from Eq. \ref{eqn:Wip1}, and by definition of CLVs (Eq. \ref{eqn:covariance})
\begin{align} 
		W^i_{x_1}
		&= \dfrac{1}{z_{x,i}}\dfrac{d}{dt} \left(\frac{d\varphi_{\curvepi}\; V^i_{\curvepi}}{z_{\curvepi, i}}\right)(0)  \\
    \notag
		&= \dfrac{1}{z_{x,i}^2} \dfrac{d}{dt}\left(d\varphi_{\curvepi}\right)(0) \; V^i_x + \dfrac{1}{z_{x,i}^2} d\varphi_x \; \dfrac{d}{dt} (V^i_{\curvepi})(0) \\
    \label{eqn:iterationStart}
		&+ V^i_{x_1} \; \dfrac{d}{dt}
		\left( \dfrac{1}{z_{\curvepi, i}}\right)(0) 
\end{align}  
By Eq. \ref{eqn:WiinNewCoordinates}, we can write the 
second term above as $(1/z_{x,i})^2\: d\varphi_x\: W^i_x.$ The first term can 
be written using the chain rule in terms of 
the $m\times m\times m$ second-order derivative
of $\varphi$, which 
is a bilinear form denoted as $d^2\varphi.$
Let the elements of the second-order derivative of the map be indexed such that 
$d^2\varphi[i,j,k] = \partial_k\partial_j\varphi_i,$ and let  
$d^2\varphi : b$ indicate the $m\times m$ matrix resulting from 
taking the dot product of the last axis of $d^2\varphi$ and the vector $b$. Then, Eq. 
\ref{eqn:iterationStart} becomes
\begin{align} 
\notag
		W^i_{x_1} 
		&= \dfrac{1}{z_{x,i}^2} d^2 \varphi_x : V^i_x \; V^i_x + \dfrac{1}{z_{x,i}^2} d\varphi_x \; W^i_x \\
    \label{eqn:iteration}
		&+ V^i_{x_1}\; \dfrac{d}{dt}
		\left( \dfrac{1}{z_{\curvepi, i}}\right)(0)
\end{align}  

\subsection{The differential CLV method: iterative orthogonal projections}
The differentiation in the third term in Eq. \ref{eqn:iteration}, carried out explicitly gives, 
\begin{align}
  \dfrac{d}{dt}
		\left( \dfrac{1}{z_{\curvepi,i}}\right)(0)  \notag
		&= -\dfrac{1}{2z_{\curvepi, i}^3} 
		\dfrac{d}{dt} \Big( (d\varphi_{\curvepi} \; V^i_{\curvepi})^T
		d\varphi_{\curvepi}\; V^i_{\curvepi} \Big) (0) \\
		\notag
		&= -\dfrac{ (d\varphi_x\; V^i_x)^T}{z_{x,i}^3} 
     \Big( d^2\varphi_x: V^i_x\; V^i_x 
     + 
    d\varphi_x\; W^i_x\Big)  \\
    \label{eqn:dz}
		&= -\dfrac{(V^i_{x_1})^T}{z_{x,i}^2}\Big(d^2\varphi: V^i_x V^i_x +
    d\varphi_x \; W^i_x \Big) . 
\end{align}
Substituting Eq. \ref{eqn:dz} into Eq. \ref{eqn:iteration},
we see that Eq. \ref{eqn:iteration} simply 
projects out the component along the $V^i_{x_1}$ direction. That is,
\begin{align} 
 \label{eqn:iterationThroughProjectionBase}
		W^i_{x_1} 
		&= \Big(I - V^i_{x_1} (V^i_{x_1})^T\Big)
		\Bigg(\dfrac{d^2 \varphi_x : V^i_x\; V^i_x + d\varphi_x \; W^i_x }{z_{x,i}^2}  
    \Bigg),
\end{align}  
where $I$ is the $m\times m$ Identity matrix. That is, the CLV self-derivatives are orthogonal to the corresponding CLVs. Before the orthogonal projection, the component along $V^i$ is given by Eq. \ref{eqn:dz}, which 
indicates the change of (the reciprocal of) the expansion factor $z_{\cdot,i}$ along $V^i$. This is a fundamental quantity that influences the unstable derivative of the conditional density of the SRB measure on the unstable manifold, and will be denoted 
$$ \alpha_{x,i} :=  \dfrac{d}{dt}
		\left( \dfrac{1}{z_{\curvepi,i}}\right)(0).$$ 
We will henceforth refer to Eq. \ref{eqn:dz} as the {\em differential expansion} equation, and see its connection to linear response in section \ref{sec:linearResponse}.  

Now, Eq. \ref{eqn:iterationThroughProjectionBase} can be 
marched forward in time recursively by replacing $W^i_{x_1}$ with 
$W^i_{x_2}$, and $W^i_{x}$ with $W^i_{x_1}$. Fixing an $x$, 
we use the subscript notation, e.g. $W_n^i := W^i_{x_n},$ 
and start from a random initial 
vector $\in \mathbb{R}^m$ as a guess for $W^i_0 := W^i_x.$ 
The following iteration is proposed as the differential 
CLV method to obtain $W^i_n$, $n\in \mathbb{Z}^+,$ $1\leq i\leq d_u$
\begin{align} 
 \label{eqn:iterationThroughProjection}
  W^i_{n+1}
    &= \Big(I - V^i_{n+1} (V^i_{n+1})^T\Big)
		\Bigg(\dfrac{(d^2 \varphi)_n : V^i_n\; V^i_n + (d\varphi)_n \; W^i_n }{z_{n,i}^2}  
    \Bigg).
\end{align} 
In Appendix section \ref{sec:AppxConvergence}, we show 
that the above equation always converges asymptotically at an exponential rate when $i=1$. For other indices $1< i\leq d_u$, the convergence 
is under certain conditions on the LEs. Thus, from here on, we restrict ourselves to chaotic attractors with one-dimensional unstable 
manifolds, where we know the differential CLV method converges asymptotically. Note that the entire procedure 
above was derived for the unstable CLV self-derivatives. For the 
stable ones, we must apply the same procedure with 
time reversal since the stable and unstable CLVs are the same, 
except their roles are exchanged upon time reversal. That is, when $d_u + 1 \leq i \leq m$, we must apply the 
above iterative procedure (Eq. \ref{eqn:iterationThroughProjection})
by replacing $\varphi$ with the inverse map, $\varphi^{-1}$. Analogously, our numerical procedure converges when using $\varphi^{-1}$, as shown in Appendix section \ref{sec:AppxConvergence}, for $i=m$ -- for the self-derivative of the most stable CLV. Finally, we remark that the differential expansion/contraction equation (Eq. \ref{eqn:dz}) is also effectively time-evolved in order to compute the projection term in the differential CLV method (Eq. \ref{eqn:iterationThroughProjection}). Thus, we obtain the scalars $\alpha_{n, i}$ along a trajectory as a byproduct.  

\section{Numerical results implementing the differential CLV method}
\label{sec:results}
In this section, we implement the differential CLV
algorithm discussed in the previous section to 
several examples of low-dimensional chaotic attractors, 
some of which were introduced in section \ref{sec:background}. In every example, the unstable 
subspace is one-dimensional (a line) and numerical 
estimates of $W^1$ are shown. The Python code for the implementation, along with the files needed to generate the plots in this section, can be found at \cite{nisha-code}.
\subsection{Validation against analytical curvature of the Solenoid map}
\label{sec:solenoid}
The Smale-Williams Solenoid map produces a well-known 
example of a uniformly hyperbolic attractor that is 
contained in a solid torus. We consider a two-parameter 
Solenoid map, which in cylindrical coordinates, is written as follows:
\begin{align}
    \varphi([r, t, z]^T) 
    = \begin{bmatrix}
        s_0 + (r - s_0)/s_1 + (\cos t)/2 \\
        2t \\
        z/s_1 + (\sin t)/2 
    \end{bmatrix}.
\end{align}
\begin{figure}
    \centering
    \includegraphics[width=\textwidth]{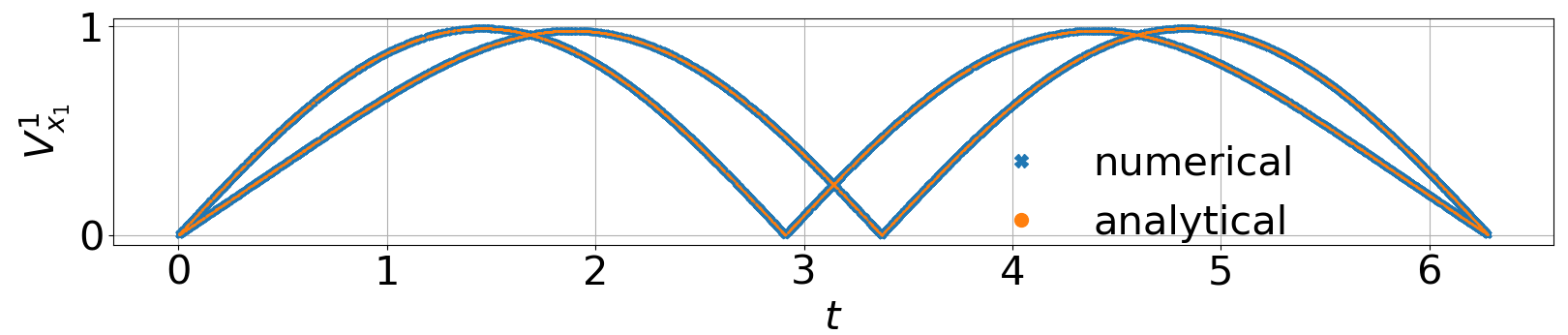}
    \includegraphics[width=\textwidth]{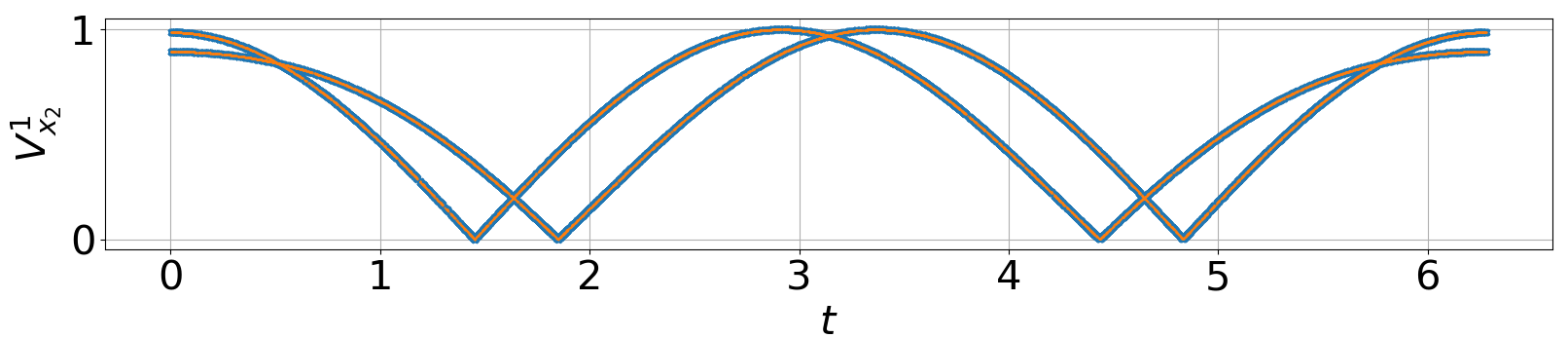}
    \includegraphics[width=\textwidth]{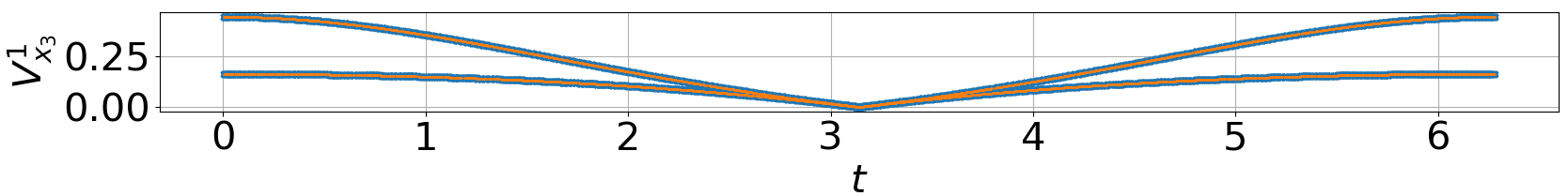}
    \caption{Comparison of the ${\rm x}_1, {\rm x}_2, {\rm x}_3$ components of $V^1$ computed analytically (orange circles) and 
    numerically (blue crosses), for the super-contracting Solenoid map}
    \label{fig:solenoid_V1}
\end{figure}
\begin{figure}
    \centering
    \includegraphics[width=\textwidth]{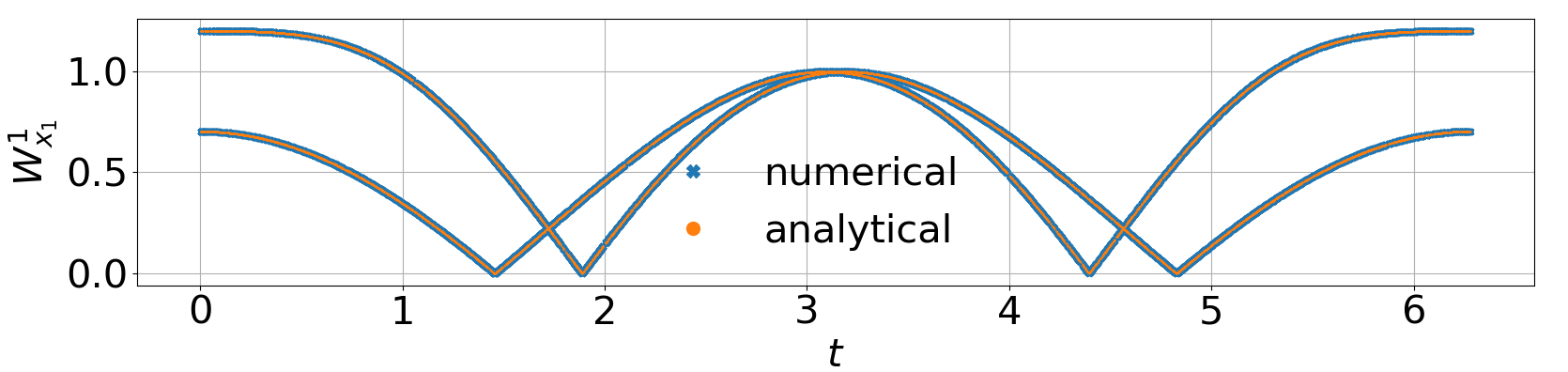}
    \includegraphics[width=\textwidth]{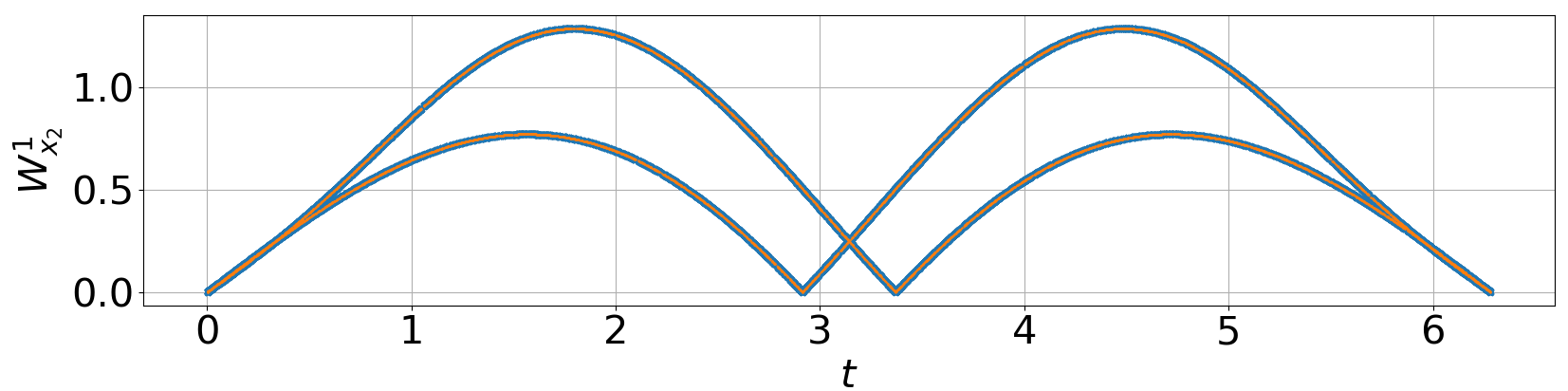}
    \includegraphics[width=\textwidth]{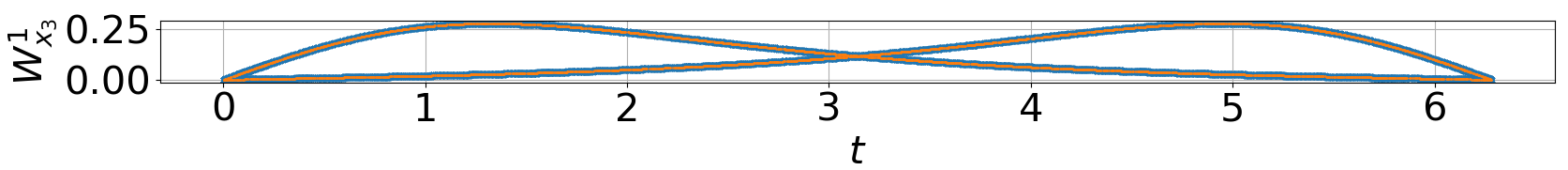}
    \caption{Comparison of the ${\rm x}_1, {\rm x}_2, {\rm x}_3$ components of $W^1$ computed analytically (orange circles) and 
    numerically (blue crosses), for the super-contracting Solenoid map.}
    \label{fig:solenoid_W1}
\end{figure}
Clearly, the parameter $s_1$ is a contraction factor along the 
$\hat{r}$ and $\hat{z}$ directions. In the limit $s_1\to \infty$, 
the attractor of the map, henceforth referred to as the super-contracting Solenoid attractor, becomes a space curve. It 
is described by the following curve parameterized by the coordinate $t$, expressed in Cartesian coordinates:
\begin{align}
	\label{eqn:solenoidAttractor}
    \gamma(t) := \begin{bmatrix}
			{\rm x}_{1,n+1} \\
    		{\rm x}_{2,n+1} \\
    		{\rm x}_{3,n+1}
    \end{bmatrix} = 
    \begin{bmatrix}
    \left(s_0 + \dfrac{\cos t}{2}\right) \cos 2t \\
    \left(s_0 + \dfrac{\cos t}{2}\right) \sin 2t \\
    \dfrac{\sin t}{2} 
    \end{bmatrix},
\end{align}
where $t = {\rm arc tan}({\rm x}_{2,n}/{\rm x}_{1,n})$. As an aside, note that in 
the $\hat{t}$ direction, the map is simply 
a linear expanding map, and hence the $\hat{t}$ 
component of the state vector has a uniform 
probability distribution in $[0,2\pi).$ We fix 
$s_0$ at 1 throughout.
The one-dimensional unstable manifold is given by 
the curve $\gamma(t)$ defined in 
Eq. \ref{eqn:solenoidAttractor}. Then, the tangent 
vector field to the curve, $\gamma'(t)$, must be along $V^1(\gamma(t))$.
This is verified numerically in Figure \ref{fig:solenoid_V1},
where the numerically computed vector field $V^1$ agrees 
closely with the unit tangent vector field 
$\gamma'(t)/\|\gamma'(t)\|$: in each of the subfigures, 
the components of the two vector fields lie superimposed 
on each other. 
\begin{figure}
	\centering
	\includegraphics[width=\textwidth]{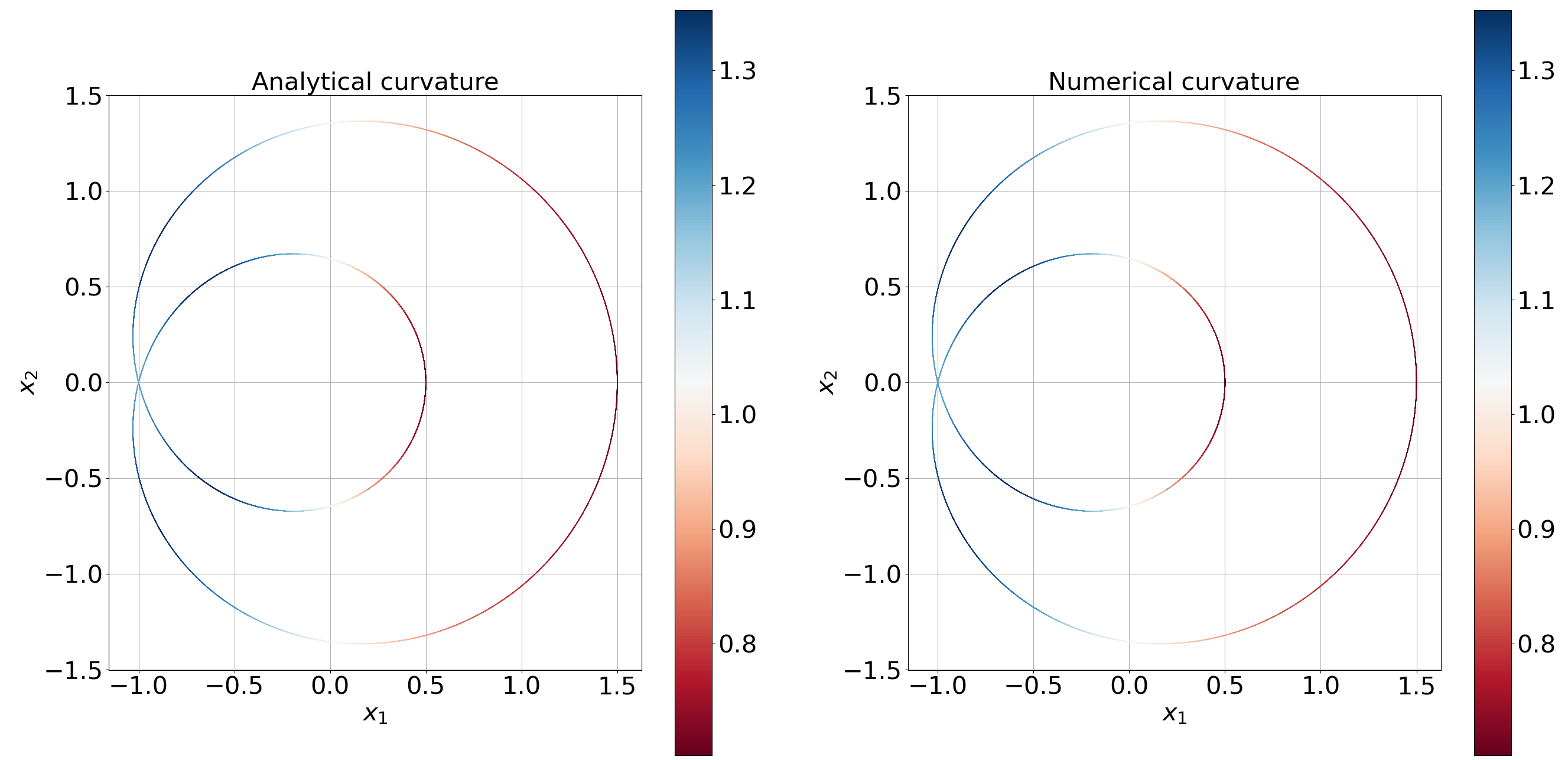}
	\caption{The vector field $V^1$ is shown for the Solenoid map. The color represents $\| W^1\|$, which is the curvature of the attractor.}
	\label{fig:curvature-solenoid}
\end{figure}
Consequently, the acceleration along the curve $\gamma(t)$, $\partial_{\gamma'(t)} \gamma'(t)$, must be in the direction of $W^1(\gamma(t))$. In particular, the acceleration in the direction 
of the unit tangent vector, $\partial_{\gamma'(t)/\|\gamma'(t)\|} (\gamma'(t)/\|\gamma'(t)\|)$, 
must match $W^1(\gamma(t))$. This is also clearly seen numerically. In Figure \ref{fig:solenoid_W1},  
each component of the two vector fields $\partial_{\gamma'/\|\gamma'\|} (\gamma'/\|\gamma'\|)$, 
computed analytically, and $W^1$, computed numerically using Eq. \ref{eqn:iterationThroughProjection},
are seen to coincide. Thus, the norms of the two vector fields are of course in close agreement as well, 
as can be seen in Figure \ref{fig:curvature-solenoid}. Both the analytically computed norm 
$\|\partial_{\gamma'/\|\gamma'\|} (\gamma'/\|\gamma'\|)\|$, and the numerically computed 
$\|W^1\|$ are shown as a colormap on the vector field $V^1 = \gamma'(t)/\|\gamma'(t)\|$. The plots in Figure \ref{fig:curvature-solenoid} are, a fortiori, a visualization of the curvature of the one-dimensional unstable manifold $\gamma(t)$. The final results of the analytical 
curvature calculations are provided in Appendix section \ref{sec:AppxSolenoid}. 

\subsection{Numerical verification of the curvature of the Lorenz attractor}
\label{sec:lorenz}
Next we consider the well-known Lorenz'63 system, given by the following system of ODEs:
\begin{align}
    \frac{d}{dt} 
    \begin{bmatrix}
			{\rm x}_1 \\
			{\rm x}_2 \\
			{\rm x}_3 
	\end{bmatrix} = F([{\rm x}_1, {\rm x}_2, {\rm x}_3]^T) := 
    \begin{bmatrix}
			10({\rm x}_2 - {\rm x}_1)\\
			{\rm x}_1(28 - {\rm x}_3) - {\rm x}_2 \\
			{\rm x}_1 {\rm x}_2 - \dfrac{8{\rm x}_3}{3}\\
    \end{bmatrix}. 
\end{align}
The map $\varphi$ is defined here to be a time-discretized form of the above system of ODEs. In particular, we use a second-order Runge-Kutta scheme with a time step of $\delta t = 0.01$. The map $\varphi(x) = x_1$, is the time-integrated solution after time $\delta t$, starting from $
x := [{\rm x}_1, {\rm x}_2, {\rm x}_3]^T \in \mathbb{R}^3.$ 
\begin{figure}
\includegraphics[width=0.58\textwidth]{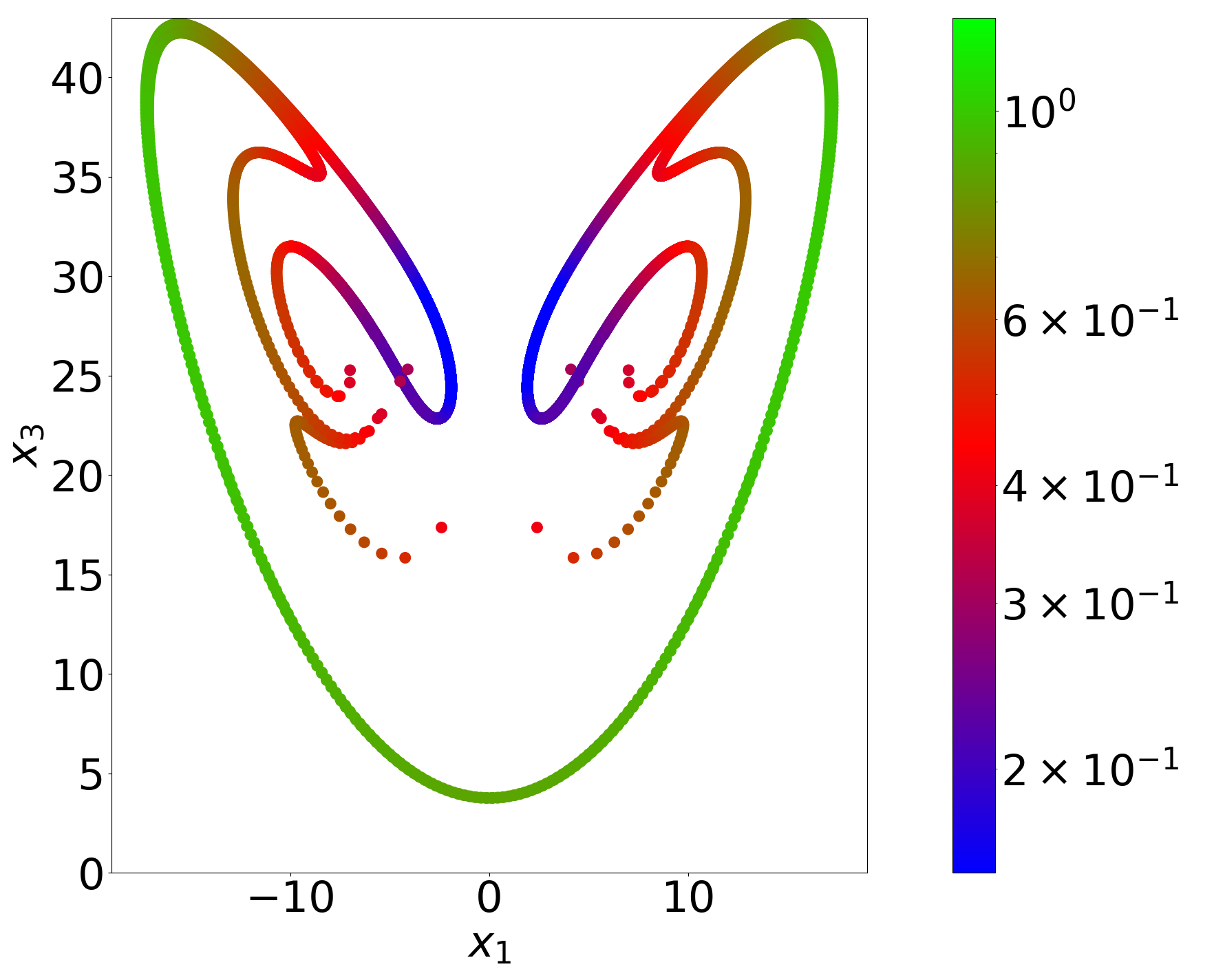}
\includegraphics[width=0.42\textwidth]{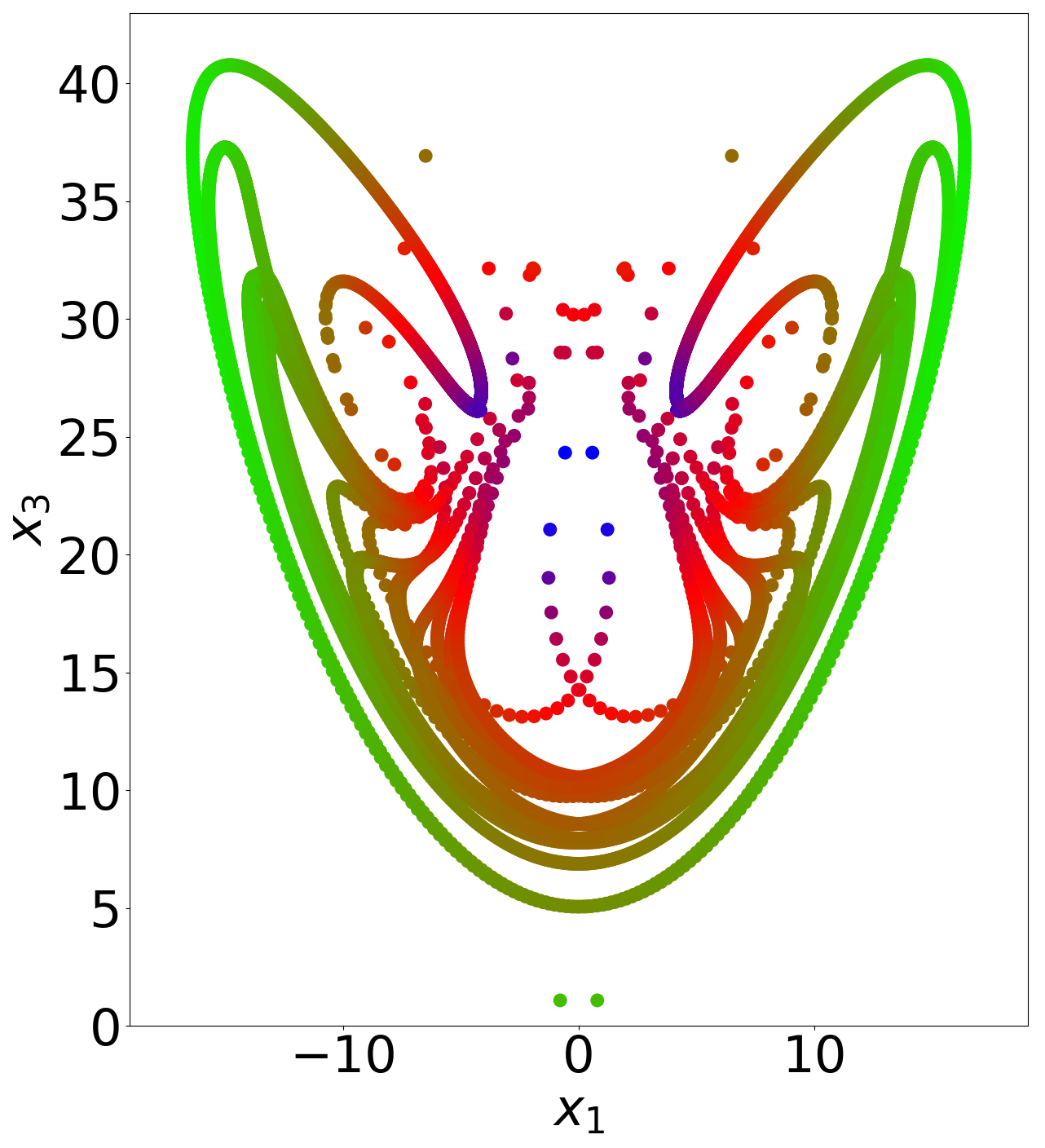}
\caption{Orbit points of the Lorenz system shown on the ${\rm x}_1$-${\rm x}_3$ plane, at $T_1=18$ (left) and at $T_2=20$ (right), colored according to the distance from their centroid normalized by the centroid $z$-coordinate. The initial conditions were 10001 equi-spaced points on the short line segment joining (-0.01,0,1) and (0.01,0,1).}
\label{fig:lorenzWu}
\end{figure}
\begin{figure}
\includegraphics[width=0.5\textwidth]{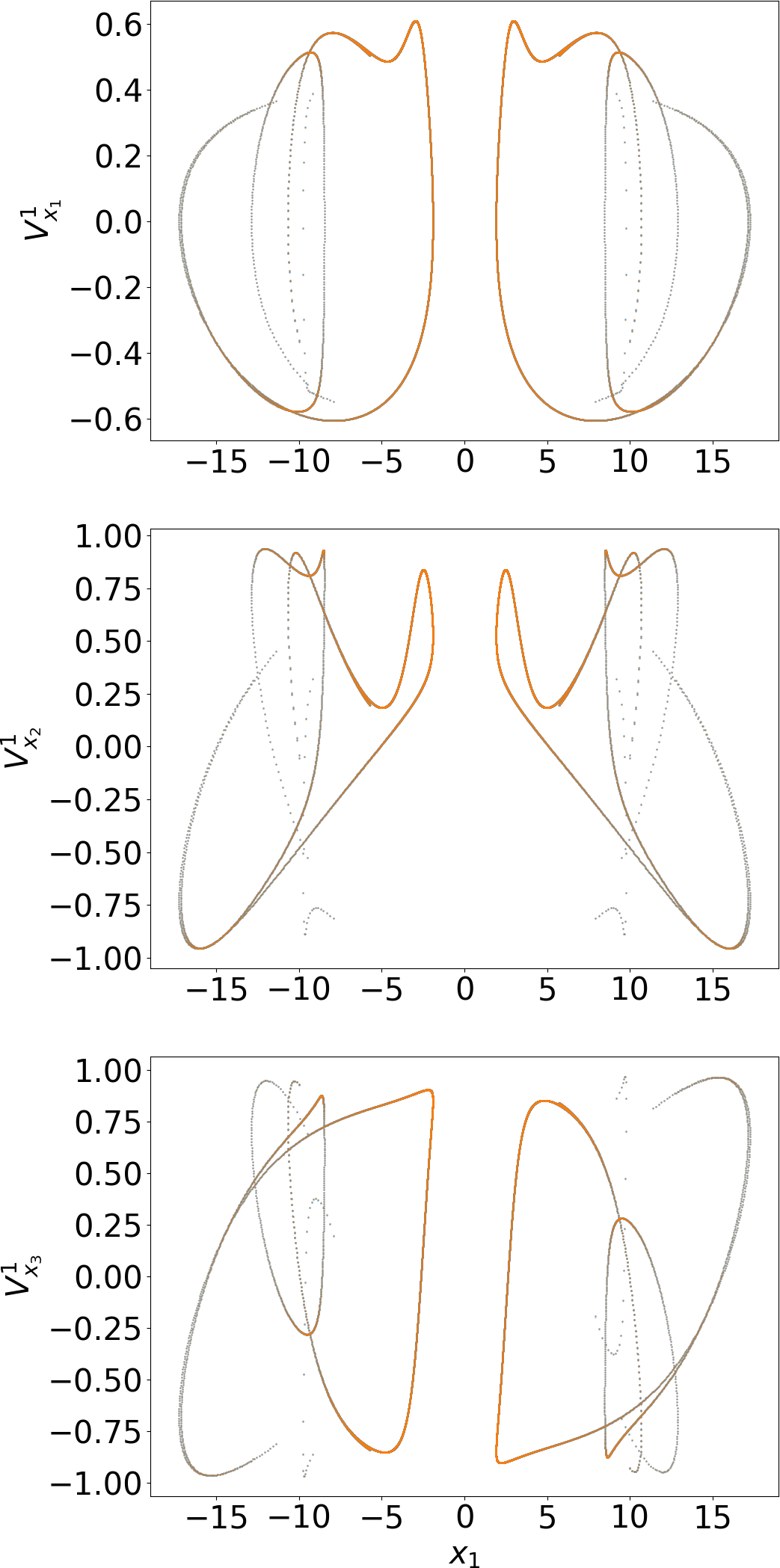}
\hspace{0.02\textwidth}
\includegraphics[width=0.5\textwidth]{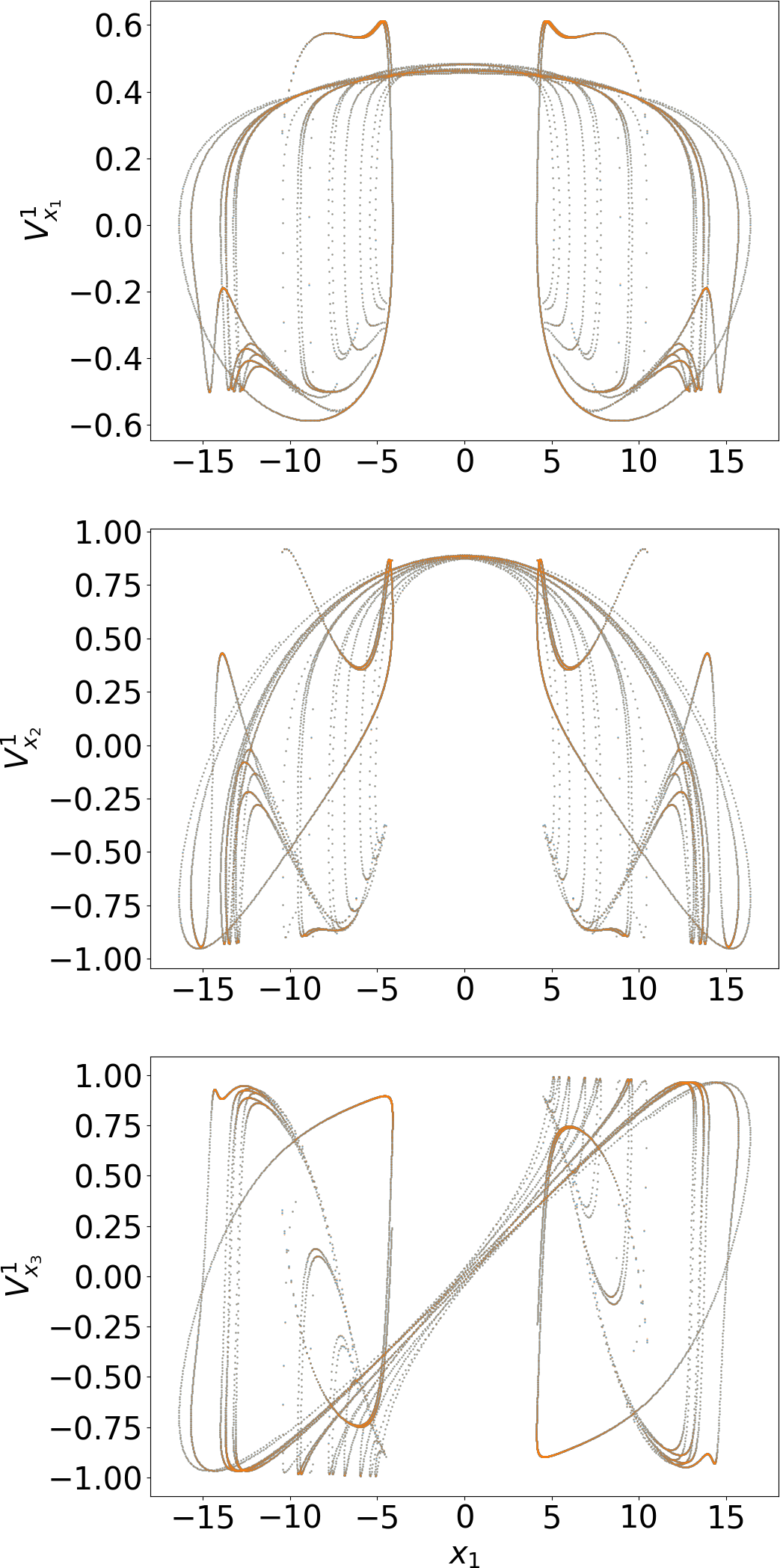}
\caption{Comparison between $V^1$ from an iteration of the tangent dynamics (shown in orange) and $V^1$ from finite difference of the primal trajectories (in blue). The first column shows the components of $V^1$ at time $T_1 = 18$ and the second column at $T_2 = 20.$ The first, second and third rows show the ${\rm x}_1$, ${\rm x}_2$, ${\rm x}_3$ components of $V^1$ respectively.}
\label{fig:lorenzV1}
\end{figure}
\begin{figure}
\includegraphics[width=0.5\textwidth]{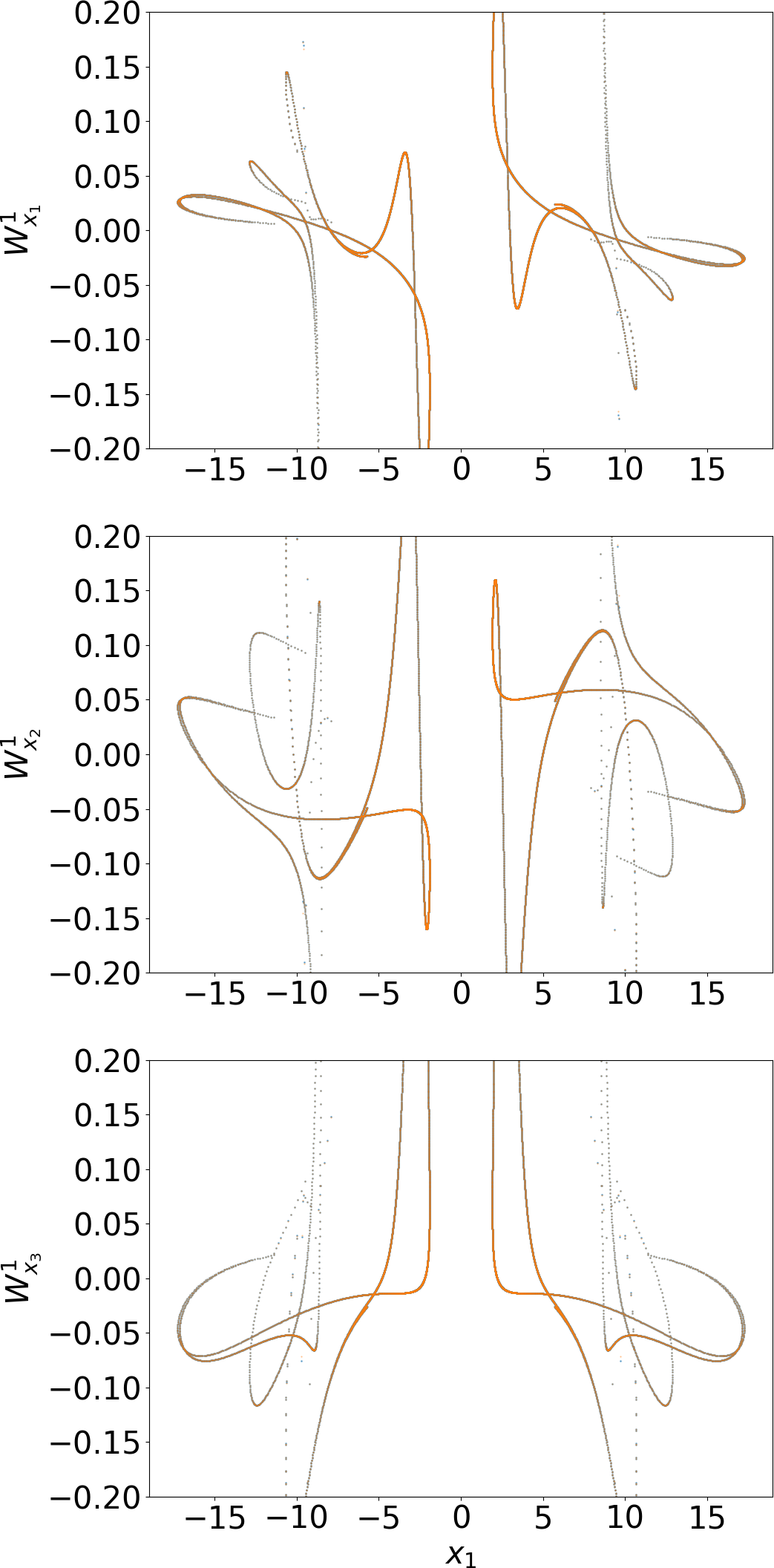}
\hspace{0.02\textwidth}
\includegraphics[width=0.5\textwidth]{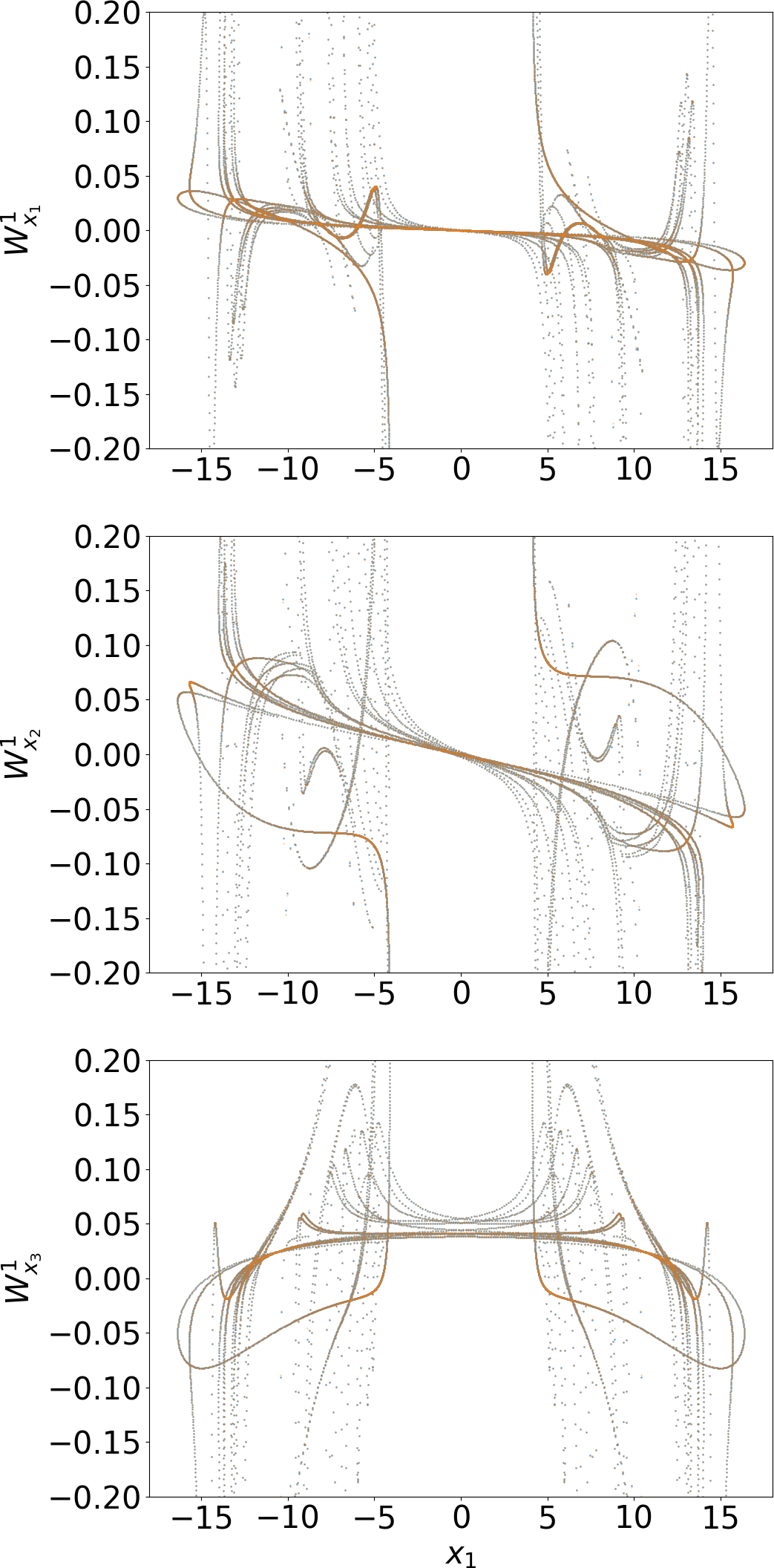}
\caption{Comparison between $W^1$ from the differential CLV method (shown in orange) and $W^1$ from finite difference (in gray). The first column shows the components of $W^1$ at time $T_1 = 18$ and the second column at $T_2 = 20.$ The first, second and third rows show the ${\rm x}_1$, 
${\rm x}_2$, ${\rm x}_3$ components of $W^1$ respectively.}
\label{fig:lorenzW1}
\end{figure}
The Lorenz'63 map defined this way has the following Lyapunov exponents: $\lambda_1 \approx 0.9$,
$\lambda_2 \approx 0$ and $\lambda_3 \approx -14.6.$ The unstable manifold, which is tangent to the CLV corresponding to $\lambda_1$, is one-dimensional. There is a one-dimensional center manifold tangent to the right hand side of the ODE, $F$. This corresponds to $\lambda_2 \approx 0$, i.e., since clearly 
$F(x_1) \approx d\varphi_x F(x),$ the tangent vector roughly parallel to $F(x) \in T_x \mathbb{R}^3$ does not show exponential growth or decay 
under the tangent dynamics. Thus, this 
map is not uniformly hyperbolic as per 
the description in section \ref{sec:uniformHyperbolicity}. 
Rather, it is a partially hyperbolic system  --a generalization of a uniformly hyperbolic system that allows a center direction -- in which the center-unstable manifold is two-dimensional and tangent to ${\rm span}\left\{F\right\} \oplus E^u$. The Lorenz attractor 
nevertheless mimics the statistical behavior 
of a uniformly hyperbolic attractor. For instance, the central limit theorem holds for H\"older continuous observables and an SRB-type invariant distribution exists \cite{araujo}.

In Figure \ref{fig:lorenzWu}, we numerically calculate the one-dimensional unstable manifold at $x := (0,0,1)$ of the Lorenz attractor. We populate 
the small line segment connecting [-0.01,0,1] and [0.01,0,1] with 10001 equi-spaced initial conditions. In Figure \ref{fig:lorenzWu}, 
these points are shown after 
time evolution for time $T_1 = 18$ or $n_1 = 1800$ steps (on 
the left) and 
$T_2 = 20$ or $n_2 = 2000$ steps (on the right).
The points that are a small distance from one another at all times up to the indicated times are considered orbits within local unstable manifolds of the reference orbit $\left\{x_n\right\}$.

Along these selected orbits, we use the following finite difference approximation to compute $V^1$:
\begin{align}
		V^1(y_n) \approx \frac{x_n - y_n}{\norm{x_n - y_n}}.
\end{align}
The 3 components $V^1_{{\rm x}_i}$, $i = 1,2,3$ 
obtained this way are shown in gray in 
Figure \ref{fig:lorenzV1}; to avoid confusing these scalar fields with $V^1(x_n)$, we do not use the shorthand notation, in this section, for $V^1(x_n)$, which refers to the first CLV at the phase point $x_n$. The scalar fields $V^1_{{\rm x}_i}$ match 
match closely the results, shown in orange, of a more typical
method of computing the first CLVs. This 
second method to compute $V^1(x_n)$ uses only the trajectory $x,x_1,\cdots,x_n$ and 
the tangent dynamics along this trajectory, and works  
as follows: randomly initialize $v(x)$ and 
propagate the tangent dynamics with repeated normalization.
\begin{align}
    v(x_{n+1}) &=  d\varphi(x_n) v(x_n),\\
    v(x_{n+1}) &\longleftarrow v(x_{n+1})/\|v(x_{n+1})\|.
\end{align}
Carrying this out for $n\in \mathbb{Z}^+$, similar to a power iteration method for the computation of the dominant eigenvector of a matrix, yields a unit vector $v(x_n)$ that aligns with $V^1(x_n).$ As confirmed in Figure \ref{fig:lorenzV1}, this procedure is equivalent to the above-mentioned finite difference procedure, as long as $y_n$ is in a small neighborhood of $x_n,$ for the length of the trajectory considered.

Having visualized $V^1$ along trajectories, 
we now compute $W^1$ using our differential CLV method
in section \ref{sec:algorithm}. To test 
its correctness, we also compute $W^1$ 
using a finite difference method as follows.
As usual, let the reference trajectory 
along which we require to compute $W^1$ 
be $x,x_1,\cdots,x_N,$ and assume that 
we know the CLVs $V^1(x), V^1(x_1), \cdots, V^1(x_N).$ Let $y, y_1, \cdots, y_N$ and 
$r, r_1, \cdots, r_N$ be two other trajectories that are at most a distance 
of ${\cal O}(1)$ away from the reference trajectory, at each of the $N$ time steps.
Then, according to our preceding discussion, 
\begin{align}
 V^1(y_n) \approx -V^1(x_n) \approx \frac{x_n - y_n}{\norm{x_n - y_n}}.
\end{align}
At each $n$, we rescale $y_n$ and $r_n$ along $V^1(x_n)$
to obtain the two points i) $\tilde{y}_n = x_n + \epsilon_{y_n} V^1(y_n)$, ii) $\tilde{r}_n = x_n + \epsilon_{r_n} V^1(r_n)$. Then, we can approximately 
compute 
$W^1(x_n)$ as 
\begin{align}
    W^1(x_n) \approx
    \frac{(\tilde{r}_n - x_n)/\epsilon_{r_n} - 
    (\tilde{y}_n - x_n)/\epsilon_{y_n}}{\norm{\tilde{r}_n - \tilde{y}_n}}.
\end{align}
In Figure \ref{fig:lorenzW1}, we plot 
the three components of $W^1$: 
$W^1_{{\rm x}_1}, W^1_{{\rm x}_2}, W^1_{{\rm x}_3}$ 
computed using the above procedure 
in gray and the same quantity computed 
using the differential CLV algorithm in 
section \ref{sec:algorithm} in orange. 
The closeness of the two results indicates 
the correctness of our algorithm. It is also a numerical verification of the fact that $V^1$ is differentiable along itself in this 
system, even though it is only partially hyperbolic.

\subsection{Qualitative verification on a perturbed cat map}
\label{sec:pcm}
\begin{figure}
	\centering
	\includegraphics[width=\textwidth]{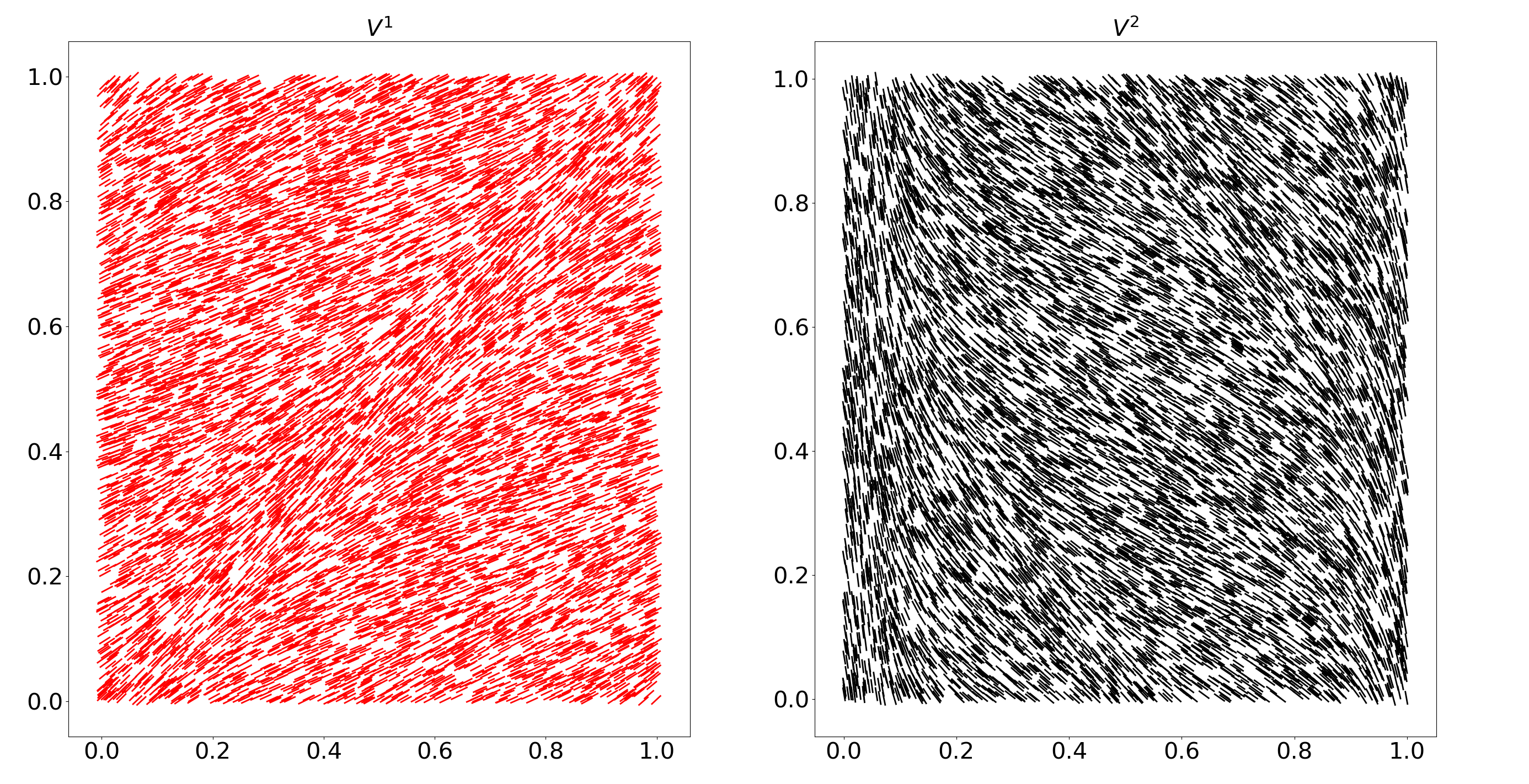}
	\caption{
	The vector fields $V^1$ (left) and $V^2$ (right) 
	are shown for the PCM at $s_1 = 0.75, s_2 = 0.2$.}
	\label{fig:V1V2-pcm}
\end{figure}

We consider a smoothly perturbed Cat map (PCM) (see 
section \ref{sec:examples}) due to Slipantschuk \emph{et al.} \cite{julia}. The PCM \cite{julia} was designed 
to be an analytic, area-preserving, uniformly hyperbolic map of the torus, whose spectral properties can be 
computed analytically. The PCM is given by
\begin{align}
		\varphi([{\rm x}_1, {\rm x}_2]) = \begin{bmatrix} 2 & 1 \\
				1 & 1 \end{bmatrix}\begin{bmatrix} {\rm x}_1 \\ {\rm x}_2 \end{bmatrix} + 
    \begin{bmatrix}
			\Psi_{s_1, s_2}({\rm x}_1) \\
			\Psi_{s_1, s_2} ({\rm x}_1) 
    \end{bmatrix},
\end{align}
where 
$$\Psi_{s_1, s_2}(y) := (1/\pi) \arctan\Big( s_1 \sin(2\pi y - s_2)/(1 - s_1  \cos(2\pi y - s_2))\Big)$$ is a perturbation whose maximum magnitude is controlled by the parameter $s_1$ and the location of the maximum, by $s_2.$ Clearly, the original Cat map is recovered at $s_1=0.$  As in the Cat map, the sum of the LEs is 0 but their values are sensitive to the parameters, with lesser sensitivity to $s_2$ when compared to $s_1$. Unlike the Cat map, the CLVs are no longer uniform in phase space and are also not orthogonal to each other. In Figure \ref{fig:V1V2-pcm}, we show the vector fields $V^1$ and $V^2$ computed at $s_1 = 0.75$ and $s_2 = 0.2.$ Notably, non-zero values of $s_1$ create a curvature in the CLVs, which is again non-uniform in space.
\begin{figure}
	\centering
	\includegraphics[width=\textwidth]{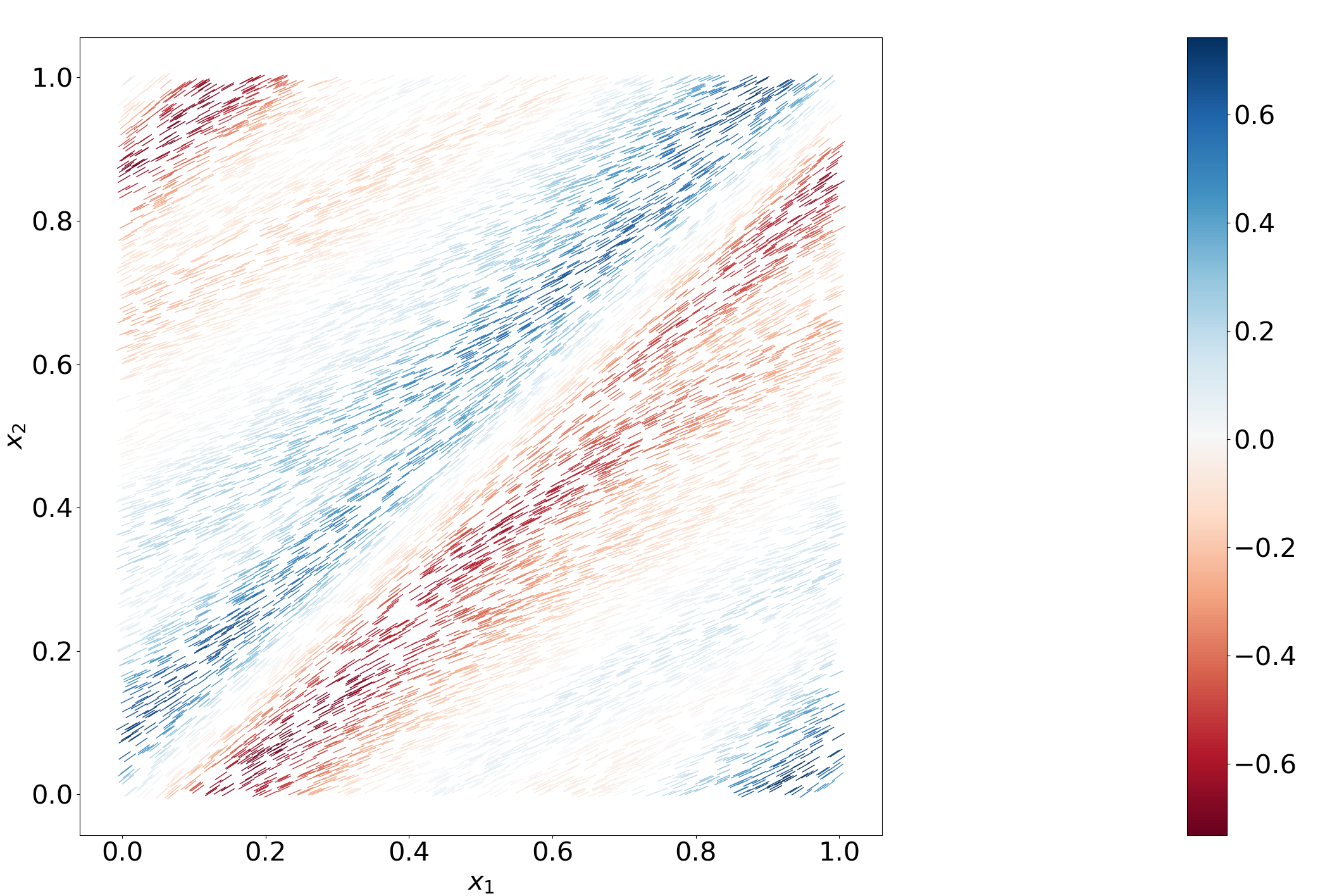}
	\caption{
	The vector field $V^1$ is shown for the PCM at $s_1= 0.75, s_2 = 0.2$. The color represents the 
	values of $\|W^1 \times V^1\|$, which equals the norm of $\| W^1\|$ multiplied by a sign representing the orientation with respect to $V^1$.}
	\label{fig:W1-pcm}
\end{figure}
We compute the self-derivative of the unstable CLV using 
our differential CLV method in section \ref{sec:algorithm}. By construction, the method produces a vector field $W^1$ that is orthogonal to $V^1$. The norm of the computed vectors, $\|W^1\|$, is shown signed according to its orientation with respect to $V^1.$ In particular, in Figure \ref{fig:W1-pcm}, we plot $\| W^1 \times V^1\|$ as a colormap on the vector field $V^1.$ Figure \ref{fig:W1-pcm} is a qualitative representation of the fact that $\norm{W^1}$ is the curvature of the unstable manifold, which is everywhere tangent to the plotted vector field $V^1.$ 
The $V^1$ self-derivative $W^1$ is the acceleration of a particle moving with the velocity field $V^1$. This 
intuitive picture is mirrored by Figure \ref{fig:W1-pcm}, 
in which $\norm{W^1}$ is higher in regions of velocity changes than where the velocity appears rather uniform (e.g. in a thin strip around the diagonal of the square). The regions of similar magnitude of acceleration but of opposite sign, reflect the symmetry in the velocity field $V^1$ about $x_1=x_2$, and moreover indicate the opposite directions of the \emph{turns} made in those regions by traveling particles.  
\subsection{Qualitative verification on the volume-decreasing perturbed Cat}
\label{sec:dpcm}
While the PCM was an example of a symplectic uniformly hyperbolic system, now we consider a dissipative uniformly hyperbolic map. We introduce another perturbed Cat map, with smooth nonlinear perturbations 
that cause the resulting map to be volume-decreasing. The norm of the 
perturbations is controlled by a set of four parameters $s = [s_0, s_1, s_2, s_3]^T$ and 
the unperturbed Cat map (the original Anosov Cat) is recovered at $s=[0,0,0,0]$. 
The map, referred to as the dissipative Cat map or DCM hereafter, is defined 
as follows:
\begin{align}
\notag
		\varphi([{\rm x}_1, {\rm x}_2]^T) &= \begin{bmatrix}
    2 & 1 \\
    1 & 1
    \end{bmatrix} \begin{bmatrix} 
			{\rm x}_1 \\
				{\rm x}_2 
    \end{bmatrix} + \left( s_0 \begin{bmatrix} 
    v_0 \\
    v_1 \\
    \end{bmatrix} + s_1  
    \begin{bmatrix}
    v_2 \\
    v_3 
    \end{bmatrix}\right)
    \sin(2\pi \tilde{V}^2\cdot x )/c  \\
    \label{eqn:dcm}
    &+ \left( s_2 \begin{bmatrix} 
    v_0 \\
    v_1 \\
    \end{bmatrix} + s_3  
    \begin{bmatrix}
    v_2 \\
    v_3 
    \end{bmatrix}\right)
    \sin(2\pi \tilde{V}^1\cdot x )/c
\end{align}
where $\tilde{V}^2 := [v_0, v_1]^T = [5,-8]^T \in \mathbb{R}^2$ is a rational 
approximation of the stable CLV of the unperturbed Cat map. Similarly, $\tilde{V}^1 := [v_2, v_3]^T = [8, 5]^T \in \mathbb{R}^2$ is a rational approximation 
of the unstable CLV of the unperturbed Cat map. The constant $c$ serves to 
normalize the perturbations and is set to $c = 2\pi(v_0^2 + v_1^2).$
\begin{figure}
    \includegraphics[width=0.5\textwidth]{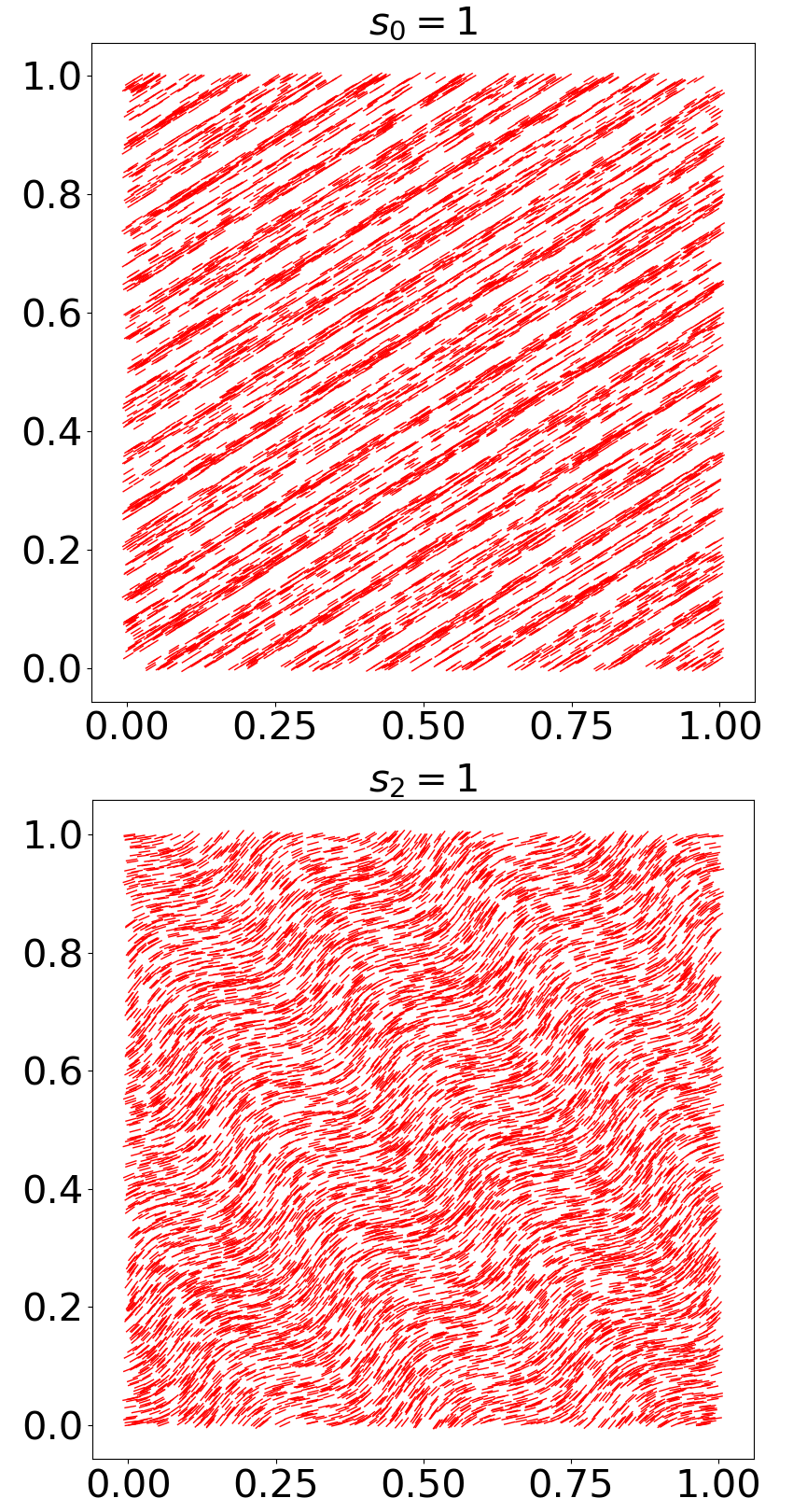}
    \includegraphics[width=0.5\textwidth]{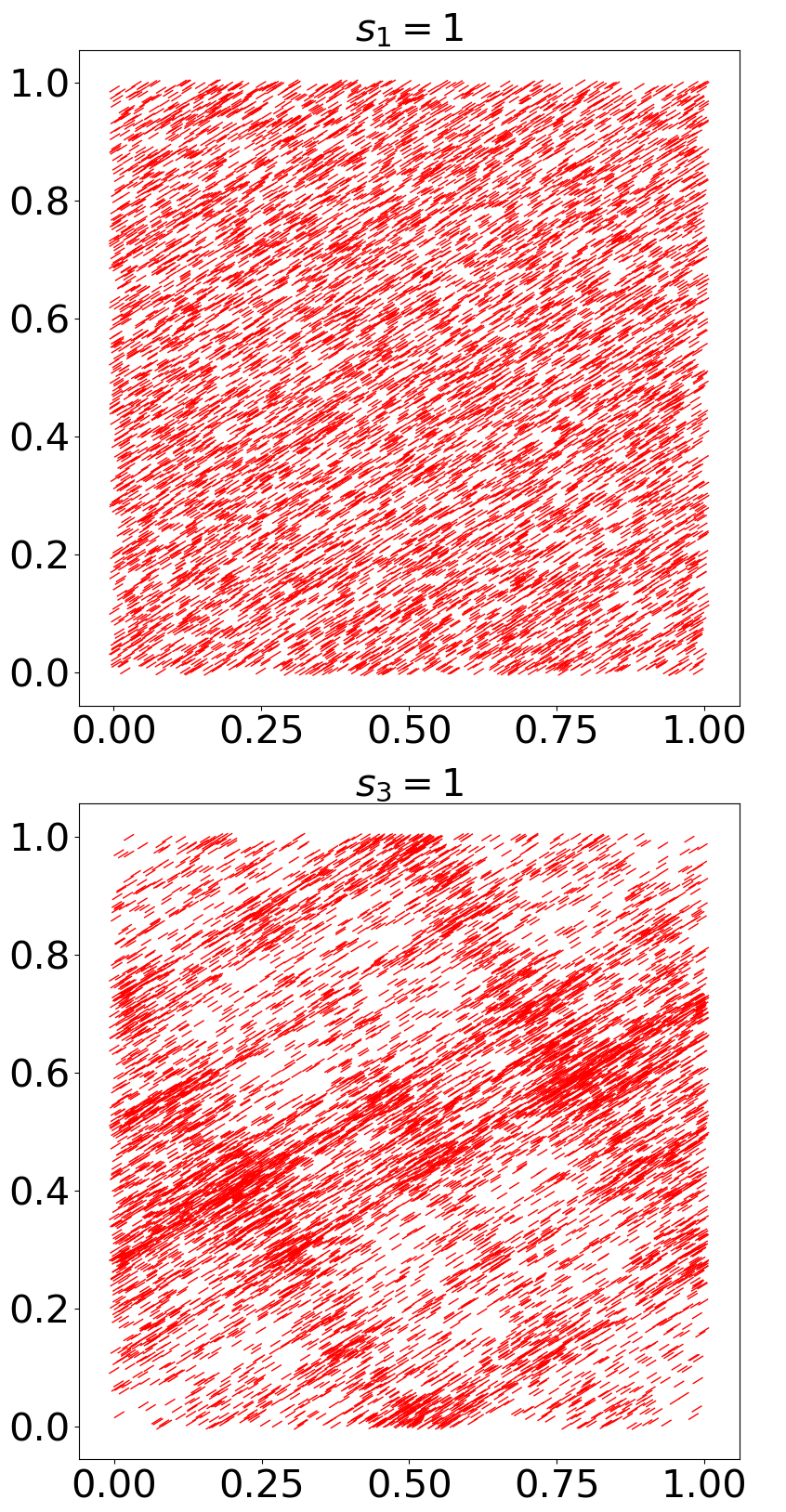}
    \caption{The vector field $V^1$ is shown for the DCM at different parameter choices. The parameters not indicated are set to 0 in each case.}
    \label{fig:DCM-V1}
\end{figure}
\begin{figure}
    \includegraphics[width=0.5\textwidth]{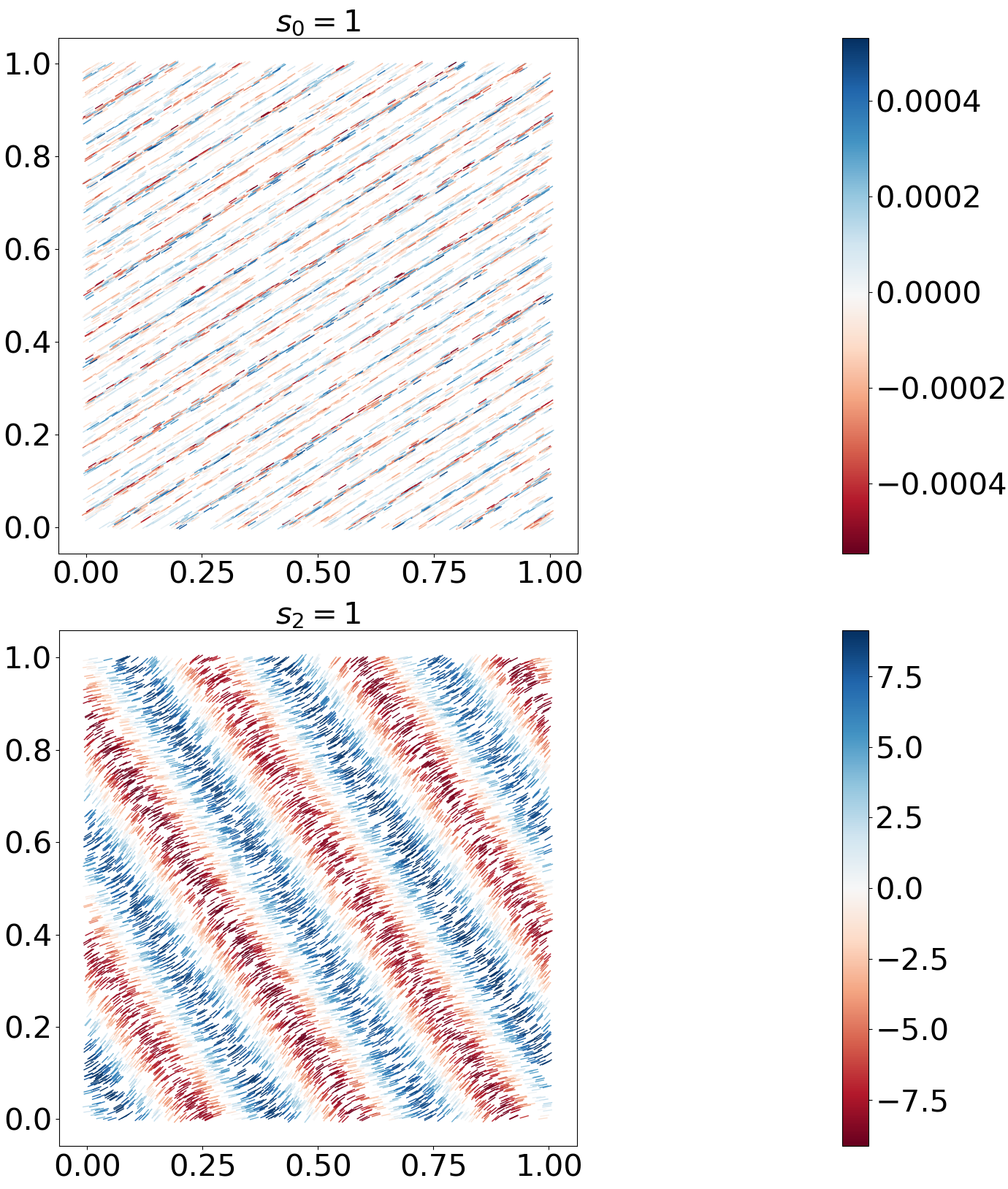}
    \includegraphics[width=0.5\textwidth]{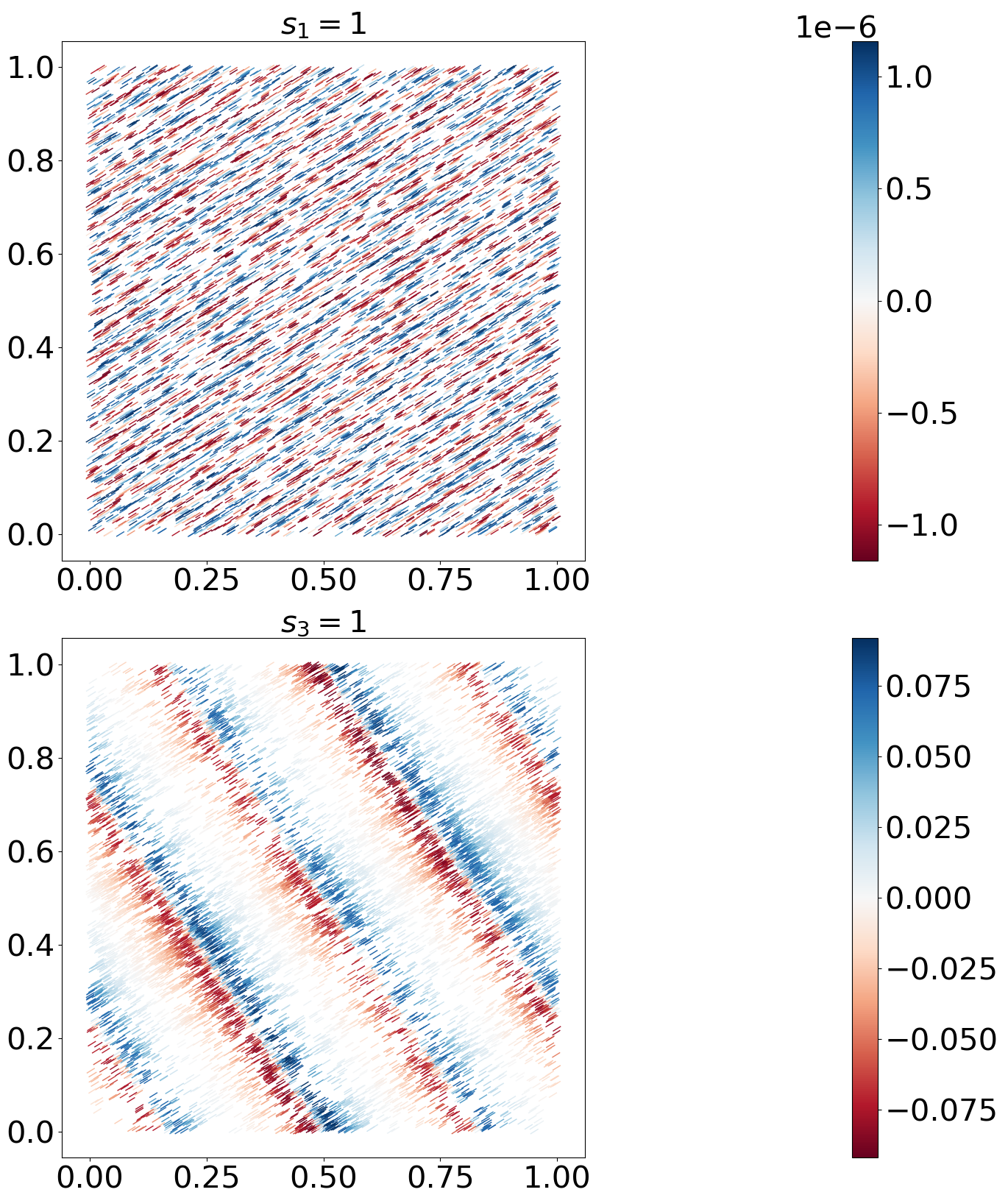}
    \caption{The vector field $V^1$ is shown for the DCM, colored according to $\norm{W^1 \times V^1}$. The parameters not indicated as 1 are set to zero in each case.}
    \label{fig:DCM-W1}
\end{figure} The four parameters together determine the norm and direction of the perturbation. In Figure \ref{fig:DCM-V1}, $V^1$ in plotted in each case of 
turning on just one of the four parameters, in order to isolate its effects. Each subfigure reflects the 
effect of a single parameter on $V^1$, in comparison to the unperturbed Cat map (in which 
$V^1$ is roughly parallel to the line $\tilde{V}^1$).
For instance, when $s = [1,0,0,0]^T$, a perturbation is applied along the 
direction $\tilde{V}^2$, which is approximately along the stable direction 
of the DCM. The norm of this perturbation varies sinusoidally with the orientation along the approximately stable direction, $\tilde{V}^2$. 
As can be seen in the top-left of Figure \ref{fig:DCM-V1}, the 
CLV $V^1$ is rather uniform in its own direction but shows a striated pattern in the perpendicular direction, roughly along $\tilde{V}^2.$
As another 
example, the bottom-left subfigure shows $V^1$ at $s = [0,0,1,0]^T.$
From Eq. \ref{eqn:dcm}, we know that $s_2$ being non-zero introduces a perturbation, along $\tilde{V}^2$, whose norm varies in the 
approximately unstable direction, $\tilde{V}^1.$ This is portrayed in the figure, wherein $V^1$ appears as waves, which are seen 
traveling approximately along $\tilde{V}^2$ but the amplitudes 
of the waves clearly vary in the perpendicular, approximately 
unstable direction. Turning on the parameter $s_1$ exchanges the roles of $\tilde{V}^1$ and $\tilde{V}^2$ when compared to when $s_2$ is non-zero. From the top-right subfigure in which the effect of $s_1$ is shown, we can see that there is no noticeable curving of the unstable manifold since the perturbation is aligned with the unstable direction.  Finally, the effect of a non-zero $s_3$ is depicted in the bottom-right of Figure \ref{fig:DCM-V1}. Here we see the compression and expansion of unstable manifolds in the unstable direction since a perturbation non-uniform in the unstable direction is applied along the unstable direction.

With this understanding of the effect of each parameter, we expect that $V^1$ would show a smaller sensitivity, in its own direction, 
when the norm of the perturbation is uniform along $\tilde{V}^1$. This is the case when $s_2, s_3$ are set to 0. 
This intuition is confirmed by the numerical results
obtained on using the differential CLV method. 
As shown in Figure \ref{fig:DCM-W1}, when either $s_0 = 1$ or 
$s_1 = 1$, and the other 3 parameters are set 
to 0, we see that the numerically computed $W^1$ has a smaller norm, 
when compared to the other cases. 

On the bottom row in Figure \ref{fig:DCM-W1} are the vector fields $W^1$ when either $s_2$ or $s_3$ are set to 1 and the rest to 0. In these cases, the norm of the perturbation varies along the approximately 
unstable direction, and this is clearly reflected in the 
higher (when compared to the other two cases) magnitudes of 
$W^1$. In addition, the variation in $W^1$ itself, which gives 
information about the second-order derivative of $V^1$, is also consistent with our expectations. For instance, 
$W^1$ shows a marked variation along $V^1$ when $s_2 = 1$ (bottom-left of Figure \ref{fig:DCM-W1}). This can be explained by the 
applied perturbations being sinusoidal in the direction of 
$\tilde{V^1}$, giving rise to a harmonic functions for the higher-order derivatives along $V^1$ as well.
Finally, when $s_3 = 1$, (bottom-right of Figure \ref{fig:DCM-W1}),
it is easy to observe that, qualitatively, the \emph{density}
of the lines $V^1$ is reflected in the magnitudes of $W^1$. 
This is not a coincidence, as we shall see in section 
\ref{sec:linearResponse}. There, we describe that $W^1$ is indirectly related to the variation in the density of the 
SRB measure on the unstable manifold, due to perturbations along 
$V^1$. Now we can see that especially  the $s_3=1$ case provides a visualization consistent with this theoretical insight. 
Particularly, the pronounced variation 
in the unstable direction (bottom-right, Figure \ref{fig:DCM-W1}), mirrors the changes in probability density on the unstable manifold, which 
is qualitatively measured by the closeness of the $V^1$ lines 
in Figure \ref{fig:DCM-V1}.

\subsection{Numerical results on the H\'enon map}
\label{sec:henon}
\begin{figure}
  \centering   
		\includegraphics[width=\textwidth]{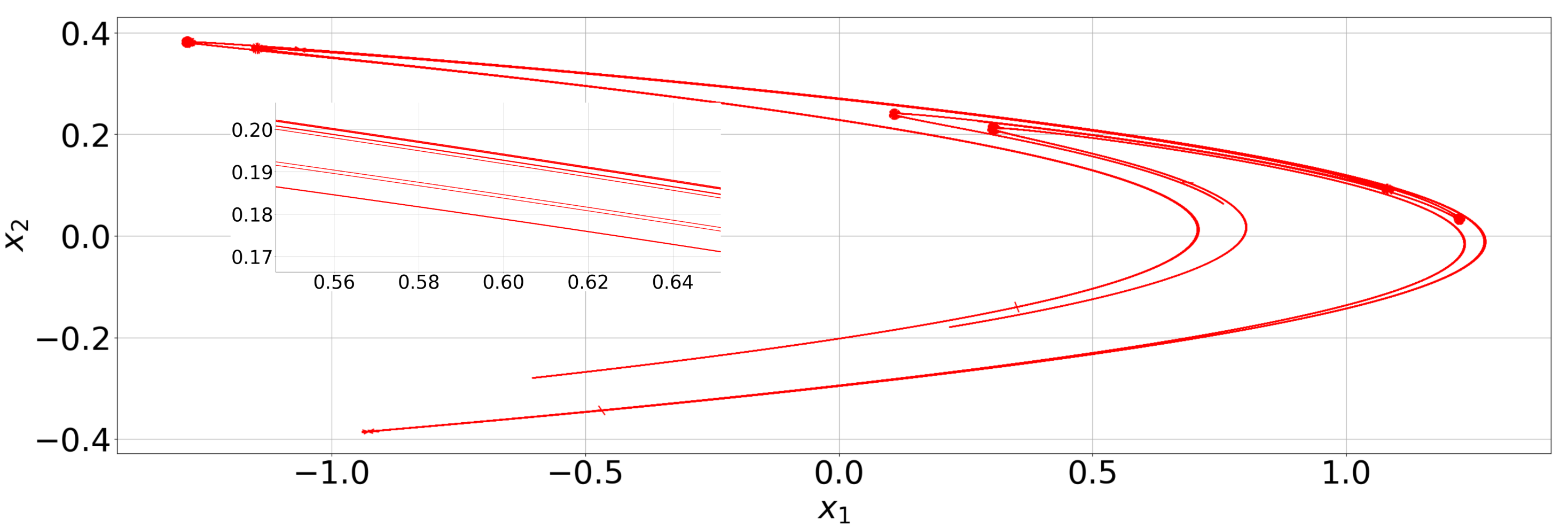}  
 \caption{The CLV $V^1$ on the henon attractor. Inset is the CLV field in a neighborhood of the fixed point $\approx (0.63,0.19).$}
  \label{fig:henonCLV}
\end{figure}

\begin{figure}
	\centering
	\includegraphics[width=\textwidth]{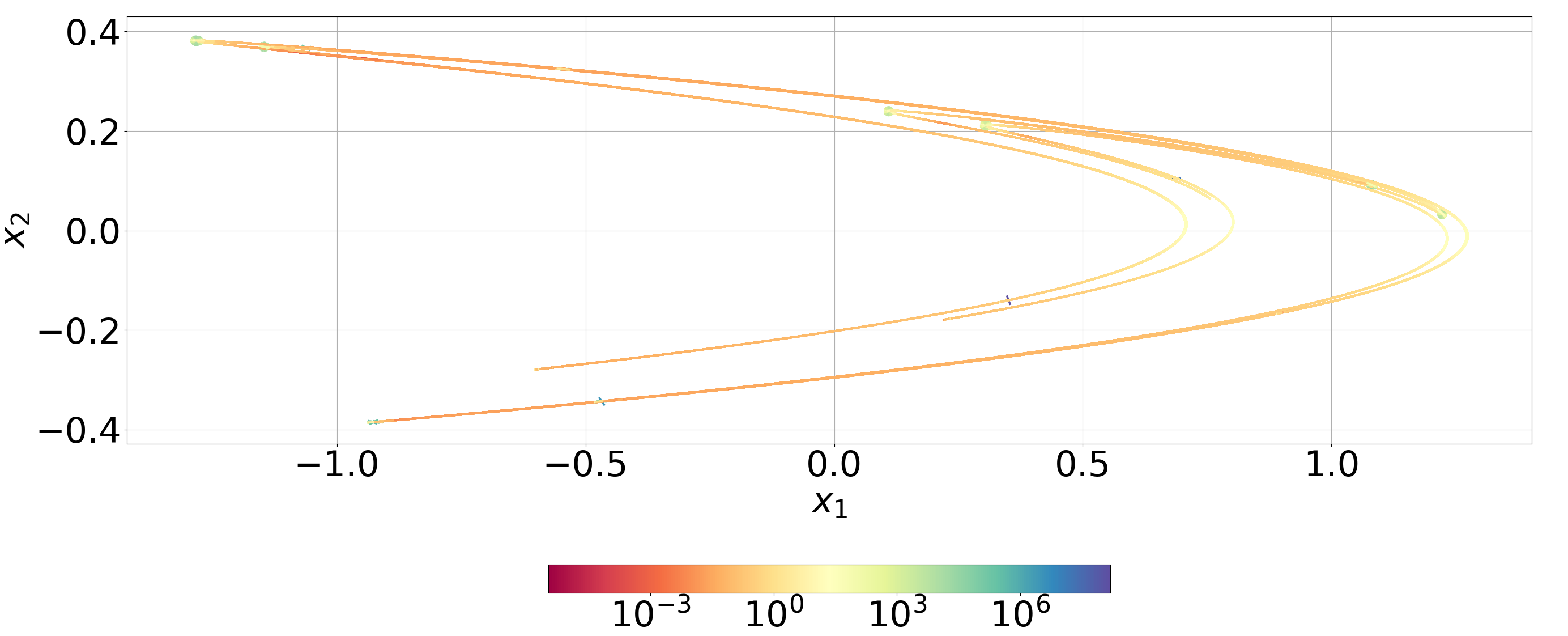}
	\caption{The vector field $V^1$ is shown for the H\'enon map. The color represents the $V^1$ self-derivative norm, $\| W^1\|.$}
	\label{fig:henonW}
\end{figure}

As our final example, we consider the classical H\'enon attractor. The 
H\'enon map is the canonical form for a two-dimensional
area-decreasing quadratic map \cite{henon}:
\begin{align}
		\varphi([{\rm x}_1,{\rm x}_2]^T) = \begin{bmatrix}
				{\rm x}_2 + 1 - s_0 {\rm x}_1^2 \\
				s_1 {\rm x}_1
    \end{bmatrix}.
\end{align}
Taking the parameters $s_0$ and $s_1$ at their standard values of $s_0 = 1.4$ and $s_1=0.3$, we obtain the H\'enon attractor, on which the CLVs are shown in Figure \ref{fig:henonCLV}. At these parameter values, the H\'enon attractor is nonhyperbolic due to the presence of tangencies between the stable and unstable manifolds \cite{henon-hyperbolicity}. On this map, we apply the differential CLV method we derived in section \ref{sec:algorithm}, and the resulting $W^1$ is shown in Figure \ref{fig:henonW}. The CLVs may not be differentiable everywhere, as seen by the large magnitudes of the numerically computed 
$W^1$ at the sharp turns in the 
attractor. 
\begin{figure}
	\includegraphics[width=0.5\textwidth]{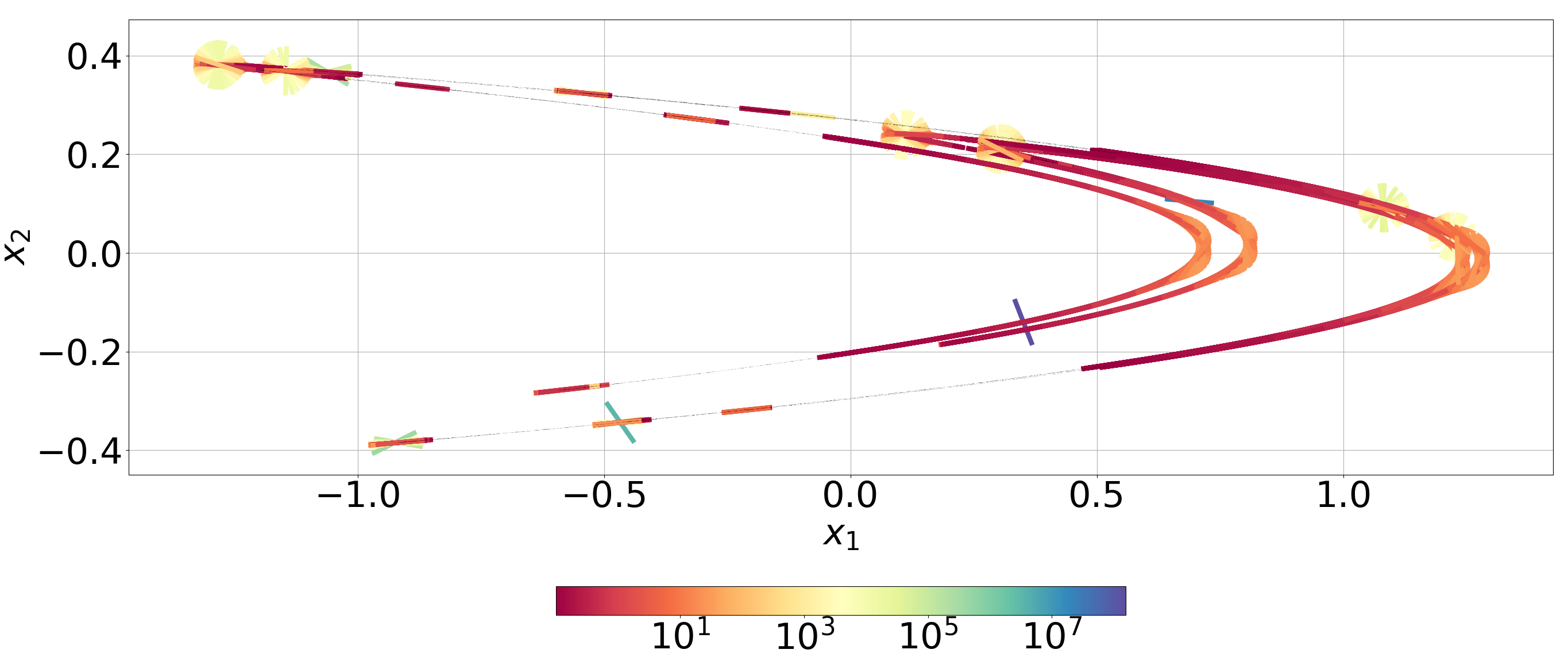}
	\includegraphics[width=0.5\textwidth]{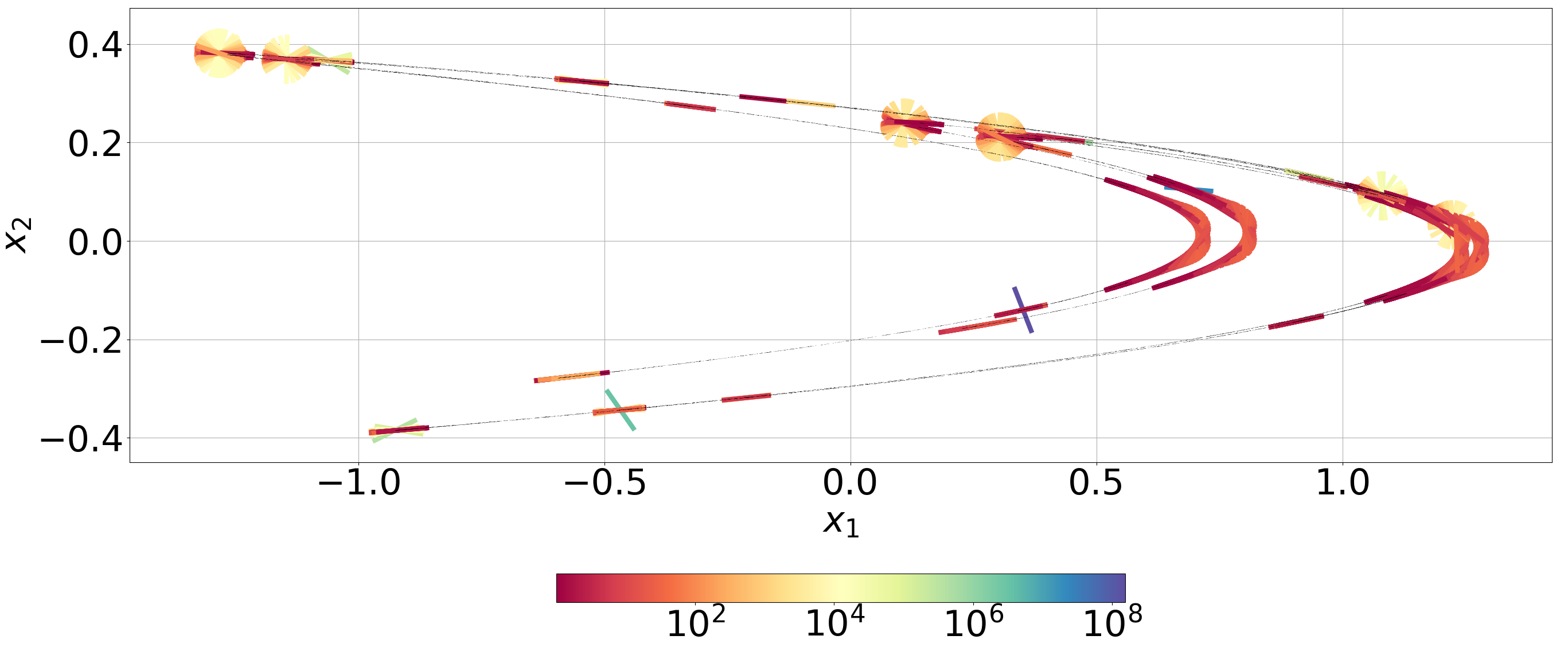}
	\includegraphics[width=0.5\textwidth]{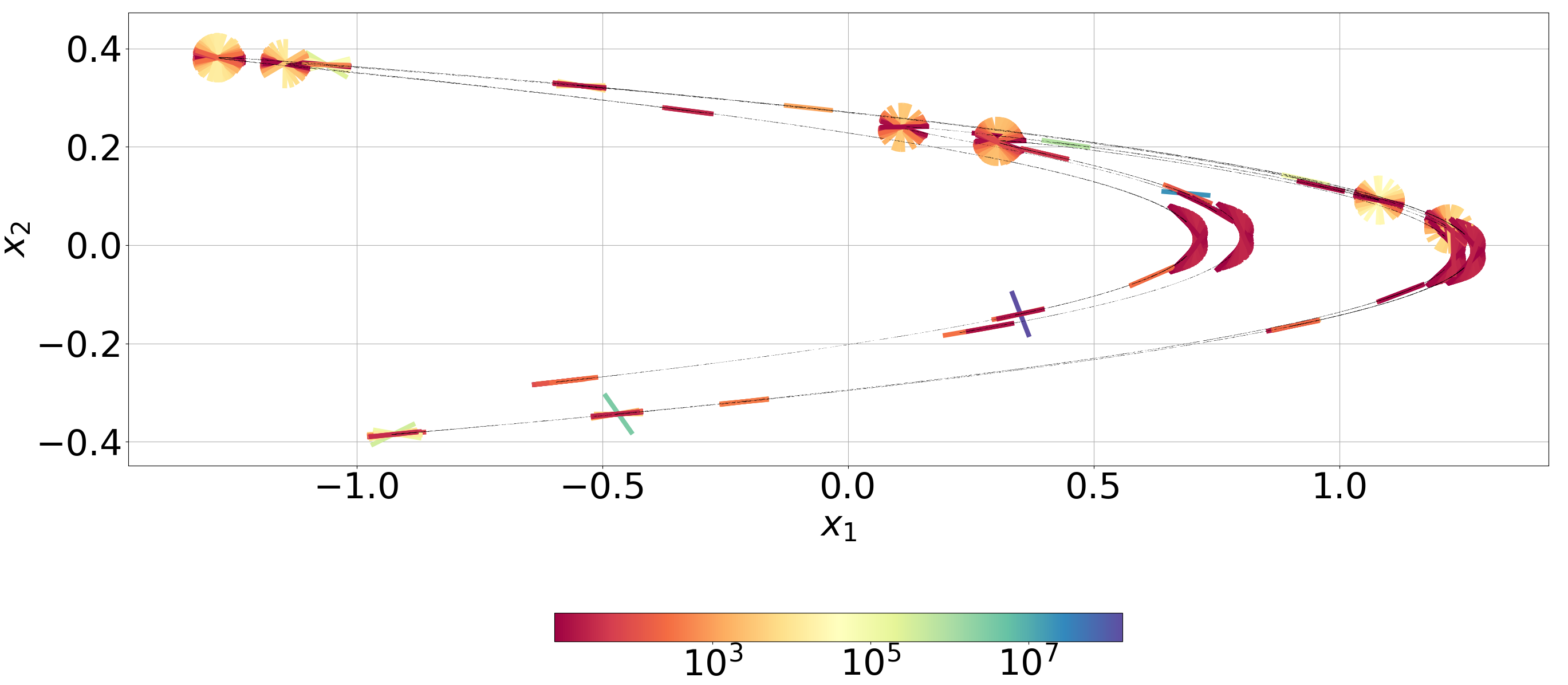}
	\includegraphics[width=0.5\textwidth]{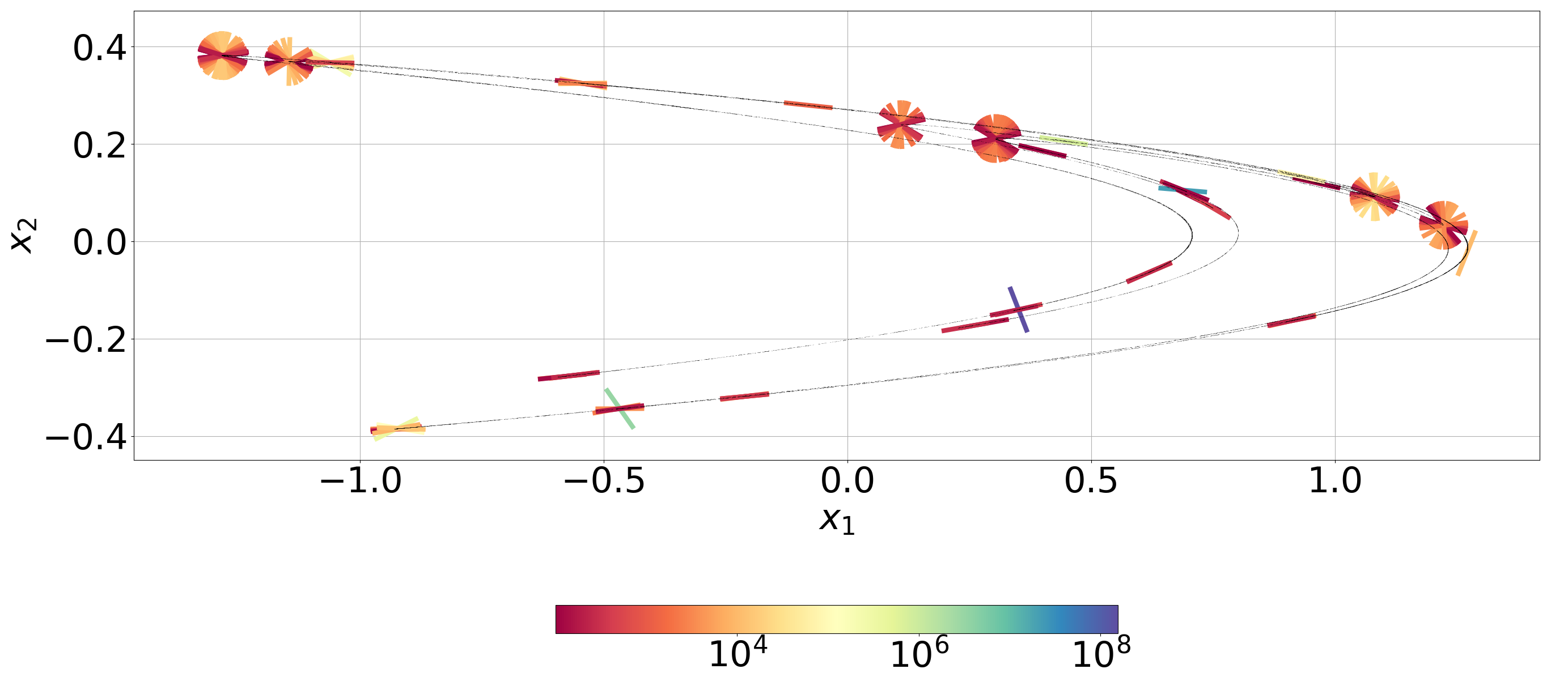}
	\caption{The vector field $V^1$ is shown for the H\'enon map. The color represents $\| W^1 \|$, the curvature of the unstable manifold.}
		\label{fig:henonW-split}
\end{figure} 
In Figure \ref{fig:henonW-split}, we dissect the derivatives further 
to investigate the issue of differentiability numerically. In each 
subfigure, the vector field $V^1$ is plotted colored according 
to $\| W^1\|;$ at the points at which $\| W^1\|$ is not in the range 
indicated by the colormap, $V^1$ is shown using thin black lines. From the top row of Figure \ref{fig:henonW-split}, it is clear that $\| W^1\| < 0.1$
for the relatively straight portions of the attractor and the points 
on the right, curved side of the attractor, still have a curvature 
less than 1. On the bottom row, the more rounded portions of the 
attractor, as expected, have a higher curvature when compared 
to the previous cases. On the bottom-right, we see that only 
the corners and turns have $\| W^1\|$ higher than 100.
Among these points, the variation in the curvature, $\| W^1\|$, 
is over six orders of magnitude, with the sharp corners having 
the highest curvatures. In this case, our numerical method for $W^1$ acts as an indicator for the lack of differentiability at some points. At least in two dimensions, this also turns out to be a detector for uniform hyperbolicity, based on our discussion in section \ref{sec:irregularity}.

\section{An application of CLV derivatives to statistical linear response}
\label{sec:linearResponse}
A landmark result in the theory of 
uniformly hyperbolic systems due to Ruelle (\cite{ruelle}\cite{ruelle1} ; \cite{gouezel} contains a modern 
proof of the result) is the smooth 
response of their statistics to parameter perturbations. Here we briefly describe this result, called the linear response formula, and draw a connection between the formula and Eq. \ref{eqn:dz}, which is the differential expansion equation. 

Consider a family of uniformly 
hyperbolic maps $\varphi_s \in C^3(\mathbb{M})$, where $s$ is a small parameter around 0. Let the reference map 
$\varphi_0$ be written simply as $\varphi$, and 
$V$ be a smooth vector field such that $\varphi_s = \varphi + s V$ up to first order in $s$. Let 
the SRB measure of $\varphi_s$ be $\mu_s$: that is, $\mu_s$ is a $\varphi_s$-invariant probability distribution on $\mathbb{M}$ such that for any continuous scalar observable $J$, the ergodic 
average starting from a $x \in \mathbb{M}$ Lebesgue-a.e., $\lim_{N\to\infty}(1/N) \sum_{n=0}^{N-1} J(\varphi_s^n(x)) = \langle J,\mu_s \rangle.$ 

Ruelle's linear response theory \cite{ruelle}\cite{ruelle1} proves the existence of the statistical response to parameter changes, $\langle J, \partial_s\mu_s\rangle$, in uniformly hyperbolic systems, including expressing this quantity as an exponentially converging series, which is known as linear response formula. The quantity $\langle J, \partial_s \mu_s\rangle$ represents the derivative with respect to $s$ of ergodic averages or equivalently ensemble averages of 
observables with respect to the SRB measure, and is of immense interest in practical applications. The statistical sensitivity $\langle J, \partial_s \mu_s\rangle$ is useful for sensitivity analysis, uncertainty quantification, model selection etc, in every scientific discipline from climate studies \cite{lucarini_climate}\cite{lucarini}
to aerodynamic fluid flows \cite{angxiu-jfm}\cite{francisco}\cite{nisha-shadowing}. The linear 
response formula \cite{ruelle}\cite{ruelle1} is as 
follows:
\begin{align}
\label{eqn:ruelle}
		\langle J, (\partial_s\mu_s)\Big|_0\rangle  &= 
		\sum_{n=0}^\infty \langle d(J\circ \varphi^n) \cdot V, \mu_0\rangle. 
\end{align}
Although the above series is exponentially converging, 
previous works \cite{nisha_ES}\cite{eyink} suggest that it is computationally infeasible to calculate the series in its original form when $V$ has a non-zero component in $E^u$, especially in high-dimensional practical systems. This is because the integrand in each term increases exponentially with $n$: $|d(J\circ\varphi^n)\cdot V| \sim {\cal O}(\exp(\lambda_1 n))$, for almost every perturbation $V$, which will have a non-zero component along $V^1$. If each term in the series is regularized by an integration by parts, the resulting form of the linear response formula is more amenable to computation. 

For a simple illustration, we consider the case of one-dimensional unstable manifolds, and fix the smooth perturbation field to be $V = a\: V^1$, which has a scalar component, $a$, along the unstable 
CLV. Applying integration by parts to Eq. \ref{eqn:ruelle} on the unstable manifold \cite{ruelle}\cite{ruelle1} (see also Appendix section \ref{sec:AppxIntByParts}), and then using the fact that ergodic averages converge to 
ensemble averages for Lebesgue-a.e. $x$,
\begin{align} 
    \label{eqn:regularizedLinearResponse}
		\langle J, (\partial_s \mu_s)\Big|_0\rangle  
    &= -\sum_{k=0}^\infty 
    \lim_{N\to\infty}\dfrac{1}{N} 
		\sum_{n=0}^{N-1} J(x_{k+n})\: (a(x_n) \: g(x_n) + b(x_n)) 
\end{align}
where 
\begin{itemize}
		\item $\rho_0$ is the density of the conditional distribution of $\mu_0$ on unstable manifolds \cite{srb};
		\item $g(x) := \dfrac{1}{\rho_0(x)} \dfrac{d (\rho_0\circ\mathcal{C}_{x,1})(t)}{dt}\Big|_{t=0},$ is the {\em logarithmic density gradient} function; and,
		\item $b(x) := \dfrac{d(a \circ\mathcal{C}_{x,1})(t)}{dt}\Big|_{t=0},$ is the derivative of $a$ along unstable manifolds.
\end{itemize}
		The computational 
infeasibility of Ruelle's original expression in Eq. \ref{eqn:ruelle} 
is overcome by Eq. \ref{eqn:regularizedLinearResponse}, as it results from regularization through integration by parts. That is, the ergodic averaging computation, listed in Eq. \ref{eqn:regularizedLinearResponse}, follows the central limit theorem, with an error convergence as ${\cal O}(1/\sqrt{N})$, when computed along an orbit of length $N$. To compute Eq. \ref{eqn:regularizedLinearResponse}, we must determine the two functions $g$ and $b$ along orbits. The derivative $b$ can be computed at any $x_n$ as $b(x_n) = (V^1_n)^T (dV)_n\: V^1_n$; to derive this expression, we use the fact that $W^1 \cdot V^1 = 0$, which follows from Eq. \ref{eqn:iterationThroughProjection}. Now, the only other unknown is the fundamental quantity -- the logarithmic density gradient denoted $g$. Using the fact that $\varphi$ preserves $\mu_0$, it can be shown that (see section 4 of \cite{nisha-s3} for an alternative derivation and \cite{adam} for an intuitive description of $g$ on one-dimensional unstable manifolds) $g$ satisfies  
 the following iterative equation along trajectories:
\begin{align}
		\label{eqn:iterationForg}
		g_{n+1} = \dfrac{g_n}{z_{n,1}} +  \alpha_{n,1}.
\end{align}
In the above equation, we use the shorthand notation $g_n := g(x_n)$ and $z_{n,1} := z_{x_n,1}$, fixing any $\mu_0$-typical $x$. Thus, Eq. \ref{eqn:iterationForg} is an iterative formula that can be used to compute $g$ along orbits. It uses the differential expansion equation (Eq. \ref{eqn:dz}) for the 
second term on the right hand side. The values of $g$ along a typical orbit, thus computed, are used in Eq. \ref{eqn:regularizedLinearResponse} to obtain the desired sensitivity.

\section{Conclusion}
\label{sec:conclusions}
In this work, we have derived a numerical method, called the differential CLV method, to compute the derivatives of Covariant Lyapunov Vectors along their own directions: the CLV self-derivatives. These directional derivatives 
exist in smooth uniformly hyperbolic systems with compact attractors. The differential CLV method converges asymptotically at an exponential rate in the case of the CLV self-derivatives corresponding to the largest and smallest Lyapunov exponents. We demonstrate the application of the differential CLV method on a variety of systems with one-dimensional unstable manifolds including a quasi-hyperbolic attractor (Lorenz'63) and 
a non-hyperbolic attractor (H\'enon). In the two-dimensional uniformly hyperbolic systems considered, 
including perturbations of the Cat map, our method provides rich visualizations of the curvature of the one-dimensional unstable manifold. A byproduct of the differential CLV method, without the orthogonal projection step (Eq. \ref{eqn:iterationThroughProjection}), known as the differential expansion equation (Eq. \ref{eqn:dz}), is fundamentally linked to the statistical linear response of a chaotic attractor. The link is through its utility to compute the divergence 
of perturbations on the unstable manifold, with respect to the SRB measure conditioned on unstable manifolds. This connection makes the differential expansion derivatives concretely useful for efficiently differentiating statistics with respect to system parameters in uniformly hyperbolic systems. The differential CLV method does not have unconditional asymptotic convergence for the self-derivatives of all CLVs, but only the most unstable and the most stable CLVs, which are treated in this work. With sufficient generalization, however, the second-order tangent equations presented in this paper can spawn applications to sensitivity analysis in chaotic systems, and beyond.

\vspace{1in}
\textbf{Acknowledgments: } We offer our sincere thanks to Dr. Jizhou Li and anonymous reviewers for helpful comments on this manuscript. 
\bibliographystyle{spmpsci}
\bibliography{refs.bib}
\appendix 
\section{The lack of differentiability of CLVs}
\label{sec:AppxContinuity}
In general, we say that a subspace $E$ is H\"older continuous 
on $\mathbb{M}$ if there exist constants $K,\delta > 0$ and $\beta \in (0,1]$ such that $\norm{E_x - E_y}_* \leq K \norm{x - y}^\beta,$
whenever $x,y \in \mathbb{M}$ are such that $\norm{x-y} \leq \beta.$
As mentioned in section \ref{sec:irregularity}, the subspaces 
$E^u$, $E^s$ are H\"older continuous 
spaces with an $\beta$ that is rarely equal to 1. The reader is referred to classical texts such 
as \cite{katok} (Chapter 19) or \cite{hasselblatt_prevalence_1999}
for a detailed exposition on H\"older structures on hyperbolic sets. 

There, the norm $\norm{\cdot}_*$ uses 
an adapted coordinate system such as the one introduced 
in section \ref{subsec:coordinateNotation}. The
set of H\"older continuous functions themselves, is independent of 
the coordinate system, however.  The norm $\norm{\cdot}_*$ used in the above references (e.g. in Theorem 19.1.6 of \cite{katok}), for our particular choice of adapted coordinates introduced in \ref{subsec:coordinateNotation}, results in the following definitions, which are exactly what one might expect. Suppose $\norm{x-y} \leq \delta$, and $Q_x, Q_y$ 
are matrix representations of the CLV basis whose 
$i$th columns respectively are $V^i_x, V^i_y$. Then, 
$\norm{E^u_x - E^u_y}_* := \norm{Q_x[:,1:d_u] - Q_y[:,1:d_u]}$
where the norm on the right hand side is a matrix norm on 
$\mathbb{R}^{m\times d_u},$ say the induced 2-norm. Here we have again used programmatic notation: given a matrix $A$, $A[:,i:j]$ refers to the columns of 
$A$ from $i$ to $j$, limits included. Similarly, for $E^s$,
$\norm{E^s_x - E^s_y}_* := \norm{Q_x[:,d_u+1:d] - 
Q_y[:,d_u+1:d]}.$ Consistent with these definitions, 
for a one-dimensional $E^i$, we have 
$\norm{E^i_x - E^i_y} := \norm{V^i_x - V^i_y},$ which 
is simply the 2-norm on $\mathbb{R}^m.$

\section{Computations on the 
super-contracting Solenoid attractor}
\label{sec:AppxSolenoid}
The super-contracting Solenoid attractor is the curve $\gamma:[0,2\pi]\to \mathbb{R}^3$ (defined in 
Eq. \ref{eqn:solenoidAttractor}) parameterized by a single parameter $t$. Since we have a closed form expression for the 
one-dimensional attractor, we can compute its tangent vector field, as:
\begin{align}
    \dfrac{d\gamma}{dt} = 
    \begin{bmatrix}
        -2 r_1(t) \sin 2t
        -(\sin t \cos 2t)/2  \\
        2 r_1(t) \cos 2t - 
        (\sin t\sin 2t)/2  \\
        \dfrac{\cos t}{2}
    \end{bmatrix},
\end{align}
where $$r_1(t) = \left(s_0 + \dfrac{\cos t}{2}\right).$$
As explained in section \ref{sec:solenoid}, $V^1(t) = \gamma'(t)/\norm{\gamma'(t)}.$ 
Further, we analytically calculate that 
\begin{align}
\notag
    \partial_{\gamma'(t)/\|\gamma'(t)\|} \left( \gamma'(t)/\|\gamma'(t)\|\right) 
    &=\dfrac{1}{2} \begin{bmatrix} 
    -(193 \cos t + 392 \cos 2t + 267 \cos 3t + 68 \cos 4t + 6 \cos 5t + 36)/c_1 \\
    -(189 \sin t + 392 \sin 2t + 267 \sin 3t + 68 \sin 4t + 6 \sin 5t)/c_1 \\
    -(19 \sin t + 8\sin t \cos t + 2 \sin t \cos 2t - 2 \sin 2t \cos t)/(c_1/2)
        \end{bmatrix} \\
        \label{eqn:analyticalSolenoidCurvature}
        &\begin{bmatrix}
        -\sin 2t/r_1 & \cos 2t/r_1 & 0
        \end{bmatrix} \; \dfrac{\gamma'(t)}{\norm{\gamma'(t)}}
\end{align}
where 
$$c_1 := 2(16 \cos t  + 2 \cos 2t + 19)^{3/2}.$$ In Figures \ref{fig:curvature-solenoid} and \ref{fig:solenoid_W1}, we observe 
that the vector field $W^1$ computed using the differential 
CLV method (Eq. \ref{eqn:iterationThroughProjection}), matches 
almost exactly against the above expression in Eq. \ref{eqn:analyticalSolenoidCurvature}. 
\section{Convergence of the differential CLV method}
\label{sec:AppxConvergence}
In this section, we show that convergence of Eq. \ref{eqn:iterationThroughProjection} is guaranteed when $i=1$. Moreover, the asymptotic convergence 
is exponentially fast. Fix
a reference trajectory $q, q_1, \cdots, $, and 
use the notation $f_n$ to denote $f(x_n)$. Let 
$W^i, W_1^i,\cdots$ and $\tildeW^i,\tildeW_1^i,\cdots$
be two sequences of vectors generated by iterating Eq. \ref{eqn:iterationThroughProjection}. Then, from Eq. \ref{eqn:iterationThroughProjection},
\begin{align}
    \norm{W_n^i - \tildeW_n^i} =
		\dfrac{1}{\prod_{m=0}^{n-1}z_{m,i}^2}\norm{
    \prod_{m=0}^{n-1} \Big( 
		(I - V^i_{m+1} (V^i_{m+1})^T)\: (d\varphi)_m
    \Big) (W^i - \tildeW^i)}.
\end{align}
We can apply Oseledets MET to the cocycle ${\rm Coc}(x_m,n) = \prod_{k=0}^{n-1}(I - V^i_{m+k+1} (V^i_{m+k+1})^T)(d\varphi)_{m+k},$ and to the Jacobian 
cocycle to obtain the following 
asymptotic inequality. In particular, using 
the relationship Eq. \ref{eqn:ergodicAverageOfzi}, we get that 
for every $\epsilon > 0$, there 
exists an $N \in \mathbb{N}$ such that for all 
$n \geq N$,
\begin{align}
\notag
    \norm{W_{n}^i - \tildeW_{n}^i} &=
  \dfrac{1}{\prod_{m=0}^{n-1}(z^i_m)^2}\norm{
    \prod_{m=0}^{n-1} \Big( 
		(I - V^i_{m+1} (V^i_{m+1})^T)(d\varphi)_m
    \Big) (W^i - \tildeW^i)} \\
    \label{eqn:asymptoticConvergence}
    &\leq e^{-2n(\lambda_i - \epsilon)}\; e^{n(\omega_i + \epsilon)} \;\norm{W^i - \tildeW^i}. 
\end{align}
In the above inequality
\ref{eqn:asymptoticConvergence}, $\omega_i := \max_{j\neq i,1\leq j\leq d_u} \lambda_j.$ Thus, 
asymptotic exponential convergence is guaranteed whenever $2\lambda_i \geq \omega_i,$ which is of course true 
when $i=1.$ 

\section{Regularization of Ruelle's formula}
\label{sec:AppxIntByParts}
Here we briefly describe the derivation of Eq. \ref{eqn:regularizedLinearResponse} from Ruelle's formula (Eq. \ref{eqn:ruelle}). The reader is referred to Ruelle's original papers \cite{ruelle}\cite{ruelle1}, or to \cite{nisha-s3} for an alternative derivation of a regularized response to unstable perturbations. In the case of one-dimensional unstable manifolds, which is the focus of this paper, we can obtain Eq. \ref{eqn:regularizedLinearResponse} by the following sequence of steps:
\begin{itemize}
		\item Disintegration of the SRB measure on the unstable manifolds. Let $\Xi$ be a partition of $\mathbb{M}$ subordinate to the unstable manifold \cite{ly}, and let $\rho_0$ be the conditional density of the SRB measure on elements of $\Xi$. Then, disintegration results in the following expression for the ($n$th term in the) linear response to the unstable perturbation $a V^1$,
\begin{align}
        \langle d(J\circ\varphi^n) \cdot a\:V^1 , \mu_0 \rangle &= \int_{\mathbb{M}/\Xi} \int_{\Xi(x)} \Big(
        d(J\circ\varphi^n) \cdot a\: V^1 \Big) \circ \mathcal{C}_{x,1}(t)\: \rho_0\circ\mathcal{C}_{x,1}(t)\: dt \: d\hat{\mu}_0(x) \\
        &= \int_{\mathbb{M}/\Xi} \int_{\Xi(x)} a\circ\mathcal{C}_{x,1}(t)
        \: \dfrac{d(J\circ\varphi^n\circ \mathcal{C}_{x,1})}{dt}(t)\: \rho_0\circ\mathcal{C}_{x,1}(t)\: dt \: d\hat{\mu}_0(x).
\end{align}
                In the above expression, $\Xi(x)$ is the element of $\Xi$ containing $x$, and the quotient measure of the SRB measure on $M/\Xi$ is denoted $\hat{\mu}_0$.
    \item Applying integration by parts on the inner integral, we obtain,
\begin{align}
        \langle d(J\circ\varphi^n) \cdot a\:V^1 , \mu_0 \rangle &= \int_{\mathbb{M}/\Xi} \int_{\Xi(x)} \dfrac{d\big((a\: \rho_0\:J\circ\varphi^n)\circ \mathcal{C}_{x,1}\big)}{dt}(t)\: dt \: d\hat{\mu}_0(x) \\
		&- \int_{\mathbb{M}/\Xi} \int_{\Xi(x)} J\circ\varphi^n\circ\mathcal{C}_{x,1}(t)\:
        \Big(
        \dfrac{a\circ\mathcal{C}_{x,1}(t)}{\rho_0\circ\mathcal{C}_{x,1}(t)}
        \dfrac{d (\rho_0\circ\mathcal{C}_{x,1})}{dt}(t) \\
        &+
        \dfrac{d (a\circ\mathcal{C}_{x,1})}{dt}(t) \Big)
        \rho_0\circ\mathcal{C}_{x,1}(t)
        \: dt\: d\hat{\mu}_0(x).
\end{align}
				The first term on the right hand side of the above equation vanishes, as noted by Ruelle \cite{ruelle}\cite{ruelle1} for arbitrary dimensional unstable manifolds in Theorem 3.1(b). Applying the divergence theorem on the first term, we obtain integrals over boundaries of the partition elements, which incur cancellations in the outer integral.
\item Using the definitions of $b$ and $g$ in the above equation, we obtain
\begin{align}
        \langle d(J\circ\varphi^n) \cdot a\:V^1 , \mu_0 \rangle &= - \langle J\circ\varphi^n
        \Big(a \: g + b \Big), \mu_0\rangle.
\end{align}
				Eq. \ref{eqn:regularizedLinearResponse} is now obtained when we rewrite the above ensemble average as an ergodic average.
\end{itemize}

\end{document}